\documentclass[amsmath,amssymb,nofootinbib]{revtex4}
\usepackage{dcolumn}
\usepackage{lipsum}
\usepackage{bm}
\usepackage{graphicx}
\usepackage{color}
\usepackage{amsmath}
\newcommand{\be}{\begin{equation}}
\newcommand{\ee}{\end{equation}}
\newcommand{\ba}{\begin{eqnarray}}
\newcommand{\ea}{\end{eqnarray}}

\newcommand{\n}[1]{\label{#1}}
\newcommand{\eq}[1]{Eq.(\ref{#1})}

\newcommand{\hhh}{\, ,\hspace{0.2cm}}

\newcommand{\tu}{\widetilde{U}}
\newcommand{\tw}{\widetilde{W}}
\newcommand{\tv}{\widetilde{V}}
\newcommand{\tnu}{\widetilde{\nu}}
\newcommand{\hu}{\widehat{U}}
\newcommand{\hw}{\widehat{W}}
\newcommand{\hx}{\widehat{X}}
\newcommand{\hv}{\widehat{V}}
\newcommand{\hnu}{\widehat{\nu}}

\newcommand{\non}{\nonumber}

\begin{document}

\title{Distorted black ring} 

\author{Shohreh Abdolrahimi$^{a}$ }
\email{sabdolrahimi@cpp.edu}
\author{Robert B. Mann$^{b}$}
\email{rbmann@uwaterloo.ca}
\author{Christos Tzounis$^{a,c,d}$}
\email{ctzounis@cpp.edu}
\affiliation{${}^a$Department of Physics and Astronomy, California State Polytechnic University, 3801 West Temple Avenue, Pomona, California 91768, USA\\
${}^b$ {Department of Physics and Astronomy, University of Waterloo, Waterloo, Ontario N2L 3G1, Canada}\\
${}^c$ {Schmid College of Science and Technology, Chapman University, 1 University Drive, Orange, California 92866, USA
}\\
${}^d$ {College of Arts and Sciences, University of La Verne, 1950 Third Street, La Verne, California 91750, USA
} }

\date{2020 March 9}

\begin{abstract}
We investigate how a static and neutral distribution of external matter distorts a five-dimensional static black ring. 
We obtain a general expression for the distorted metric in terms of the background metric functions and the distortion fields, and find a multipole expansion for the latter.  
We demonstrate that the gravitational field of these external sources can be adjusted to remove the conical singularity of the undistorted black ring solution. We analyze  properties of the distorted black ring for the specific cases of dipole and quadrupole distortions.
\end{abstract}

\maketitle

\section{Introduction}
Black holes and black objects interact with external matter and fields. In order to gain a full understanding of the general theory of relativity and the properties of black holes as gravitational objects predicted by the theory of general relativity, one needs to study the interaction of black holes with matter and sources. There are several ways to study interacting black holes. Although the best way of studying such dynamical systems is numerical analysis, considerable insight can be gained from studying exact (or approximate) solutions describing a black hole tidally distorted by external matter. {\it Distorted black holes} have been constructed and studied by many  authors \cite{SCN1,dis1,dis2,dis3,dis4,dis5,dis6,SCN3,dis7,dis8,Frolov:2007xi,SDF,13s,Poisson:2009qj,Ansorg:2010ru,Abdolrahimi:2013cza,Abdolrahimi:2014qja,Abdolrahimi:2015rua,Abdolrahimi:2015gea,Abdolrahimi:2015kma,Shoom:2015rda,Semerak:2016gfz,Shoom:2015slu,Abdolrahimi:2014msa,Basovnik:2016awa,Abdolrahimi:2017pmt,Kunz:2017mfe,Grover:2018tbq}. A distorted black hole can approximate a dynamical black hole that relaxes on a timescale much shorter than that of the external matter.  

Originally, the term ``distorted black holes'' was used to describe an asymptotically flat black hole solution possessing higher mass multipole moments. Such is, for example, the Erez and Rosen solution \cite{SCN1}, that represents a generalization of the Schwarzschild black hole with a quadrupole moment. However, by Israel's theorem \cite{SCN2}, these solutions suffer from the appearance of curvature singularities on the horizon or in its vicinity. The term ``distorted black hole'' then broadened to also refer to an asymptotically nonflat black hole solution that is considered to be valid only locally in a certain neighborhood of the black hole horizon. Such solutions are interpreted as describing a black hole located in the gravitational field of external sources. Although the matter sources are not explicitly included, the solution contains information about their influence on the black hole properties. 

We consider only this class of solutions in this paper. This has the major advantage that these solutions are valid for broad classes of external matter, the only restrictions coming from some regularity conditions,  and any symmetry imposed on the solution. One of the first solutions of this class was constructed in 1965 by Doroshkevich, Zel'dovich, and Novikov \cite{SCN3}, who considered the Schwarzschild black hole in an external quadrupole gravitational field. Chandrasekhar obtained the equilibrium condition for a black hole in a static external gravitational field \cite{dis3}. Geroch and Hartle \cite{dis1} considered general static black holes in four dimensions in the presence of external matter fields, and performed a fundamental analysis of the global characteristics of these solutions. 

In higher dimensions, there is a menagerie of other black hole solutions that do not have spherical horizons, such as black rings, black saturns, black helical rings, di-rings, and bicycling black rings, as well as more general blackfolds \cite{ERW,ER2,Elvang:2007rd,Iguchi,Evslin:2007fv,Elvang2008,Izumi2008,Emparan:2011br}. While the list may not be complete, it is necessary to investigate the features of distorted versions of these black objects to investigate fully which properties of black holes or black objects are more universal. 

The construction of the vacuum solution of the Einstein equations representing a static black hole in a static external axisymmetric gravitational field relies on the {\it Weyl form} \cite{WeylS}. 
A method for the construction of distorted higher-dimensional black holes/objects based on the {\it generalized Weyl form} \cite{ERW} was developed \cite{13s}. 

A four-dimensional Weyl solution is characterized by two orthogonal commuting Killing vector fields generating an $R^1\times U(1)$ isometry group. The generalized Weyl solution (in a d-dimensional spacetime, $d>4$) is characterized by $d-2$ orthogonal commuting Killing vector fields. In the generalized Weyl form, the general solution of the $d$-dimensional vacuum Einstein equations that admits $d-2$ orthogonal commuting non-null Killing vector fields is given in terms of $d-3$ independent axisymmetric solutions of Laplace's equation in three-dimensional flat space. The generalized Weyl solution allows for the construction of many interesting black objects of different horizon topology and configuration (see, e.g., \cite{24SAch, ERW}). The first distorted solution constructed by this method was a distorted five-dimensional  Schwarzschild-Tangherlini black hole \cite{13s}. Subsequently, a solution representing a distorted five-dimensional Reissner-Nordstr\"om black hole was constructed \cite{Abdolrahimi:2013cza}. A new exact solution of the 5D Einstein equations in vacuum describing a distorted Myers-Perry black hole with a single angular momentum was obtained \cite{Abdolrahimi:2014qja}. Locally, the solution is interpreted as a black hole distorted by a stationary $U(1)\times U(1)$ symmetric distribution of external matter. In this paper, we construct a distorted black ring solution using the generalized Weyl solution. 

The study of distorted black holes/objects is important from a theoretical viewpoint. They are more general stationary and axisymmetric solutions than isolated black holes, and they can provide deeper insights into black hole properties. A series of studies were devoted to investigating how the properties of isolated black holes are influenced if they are distorted by an external matter field, and which of them remain unaffected. It was established that the 4D static distorted black holes belong to the Petrov type D class on the horizon, like their asymptotically flat counterparts, although in the rest of the spacetime they are algebraically general \cite{Papadopoulos1984}. Within the framework of isolated horizons, it was proven that a local first law of thermodynamics is valid on a distorted black hole's horizon \cite{Ashtekar1999, Ashtekar2000,Ashtekar2000v2}, which possesses the same form as the first law for the corresponding asymptotically flat black holes.  On the other hand, the analysis of  distorted black holes demonstrates that some of the other features of  black holes are not as universal. For example, it was demonstrated that, in the case of a distorted five-dimensional Myers-Perry black hole, the ratio of the horizon angular momentum and the mass $J^2/M^3$ is unbounded, and can grow arbitrarily large \cite{Abdolrahimi:2014qja}. Similarly, for a distorted Kerr black hole, the solution is regular outside the horizon even though the spin parameter can satisfy $J^2/M^4>1$ \cite{Abdolrahimi:2015gea}. This is in contrast to isolated black holes, where such ratios of angular momentum and mass lead to a naked singularity. Studies of the local shadow of the distorted Schwarzschild black hole shows that the external matter sources modify the light ring structure and lead to the appearance of multiple shadow images \cite{Abdolrahimi:2015kma,Grover:2018tbq}.

Classical general relativity in more than four spacetime dimensions is interesting to study as an extension of Einstein's theory, and in particular its black hole solutions, for at least the following reasons: String theory,  TeV-scale gravity, and brane models require more than four dimensions. The AdS/CFT correspondence relates the properties of a $d$-dimensional black hole with those of a quantum field theory in $d-1$ dimensions. As mathematical objects, black hole spacetimes are among the most important Lorentzian Ricci-flat manifolds in any dimension (see \cite{ER3}). In addition to these applications of the subject, there exists intrinsic interest in higher-dimensional gravity. 
A number of classical theorems show that black holes in four spacetime dimensions are highly constrained objects. For a stationary, asymptotically flat, vacuum black hole, event horizons of nonspherical topology are forbidden \cite{HAW}. In five dimensions, the situation is not so simple. 
$S^1\times S^2$ is one of the few possible topologies for the event horizon in five dimensions. 
An asymptotically flat, stationary, vacuum solution with a horizon of topology $S^1\times S^2$, a rotating black ring was constructed \cite{ER2}. An uncharged static black ring solution is presented in \cite{ERW}, but it contains conical singularities. We construct a solution representing a local distorted uncharged static black ring distorted by external static and neutral distribution of external matter which can be free of conical singularities. 

Our paper is organized as follows: In Sec. II, we discuss the generalized Weyl form.
In Sec. III, we overview the black ring solution. In Sec. IV we construct the metric of a distorted black ring. In Sec. V, we analyze the general properties of the spacetime and conditions on the distortion fields. In Section VI, we analyze the spacetime properties of the distorted black ring further for dipole and quadropole distortions. In this paper, we use the following convention of units:
$G_{(5)}= c = 1$, the spacetime signature is $+3$, and the
sign conventions are those adopted in \cite{MTW}.

\section{Generalized Weyl solution}
In four dimensions a general static, axisymmetric solution of vacuum Einstein equations can be presented in the Weyl form \cite{WeylSAD}. In this form, the vacuum Einstein equations simplify. One of these equations is a three-dimensional Laplace equation defined in a flat auxiliary space which is solved by the first of the metric functions. The second metric function can be found by a simple line integral defined in terms of the first one. This simple structure of the Einstein equations allows one to find exact analytical solutions to many interesting models of classical general relativity, e.g., the Israel-Khan solution representing a set of collinear Schwarzschild black holes \cite{IsKh}, a black hole with a toroidal horizon \cite{dis2}, a four-dimensional compactified black hole \cite{FF}, and a distorted black hole \cite{Isra:73,dis3,Chandrabook,Frolov:2007xi}. The four-dimensional Weyl solution admits an isometry group $\mathbb{R}^1_t\times O(2)$. In other words, the Weyl solution is characterized by two orthogonal, commuting Killing vectors $\xi^\alpha_{(t)}=\delta^\alpha_{\,\,\, t}$ and $\xi^\alpha_{(\phi)}=\delta^\alpha_{\,\,\, \phi}$, which are generators of time translations and two-dimensional rotations about symmetry axis, respectively. 

The $d$-dimensional generalization of the Weyl solution in vacuum admits $d-2$ commuting, non-null, orthogonal Killing vector fields (see \cite{ERW} and \cite{ER}). Note that the static and axisymmetric generalization of the Weyl form that admits the isometry group $\mathbb{R}^1_t\times O(d-2)$ is not known (see, e.g., \cite{May}).

As with the four-dimensional Weyl solution,  the generalized Weyl solution is defined by metric functions that solve the corresponding Laplace equation. According to analysis presented in \cite{ERW}, there are two classes of the generalized Weyl solution. The first class is characterized by $d-3$ metric functions that solve the three-dimensional Laplace equation and the remaining metric function is defined by a simple line integral of them. The second class is characterized by $d-2$ metric functions that solve the two-dimensional Laplace equation. This class has no four-dimensional analogue.  Here we discuss a five-dimensional Weyl solution of the first class, which is characterized by three commuting, orthogonal Killing vector fields one of which $\xi^\alpha_{(t)}=\delta^\alpha_{\,\,\, t}$ is timelike, and the other two $\xi^\alpha_{(\chi)}=\delta^\alpha_{\,\,\, \chi}$ and $\xi^\alpha_{(\phi)}=\delta^\alpha_{\,\,\, \phi}$ are spacelike. These Killing vectors are generators of the isometry group $\mathbb{R}^1_t\times U_\chi(1)\times U_\phi(1)$. Thus, the five-dimensional Weyl solution can be presented as follows:
\ba\n{1.1}
ds^2&=&-e^{2U_1}dt^2+e^{2\nu}(dz^2+d\rho^2)+e^{2U_2}d\xi^2+e^{2U_3}d\phi^2\,~,\non\\
\ea
where $t,z\in(-\infty,\infty)$, $\rho\in(0,\infty)$, and $\chi,\phi\in[0,2\pi)$ are Killing coordinates. The metric functions $U_i$, $i=1,2,3$, and $\nu$ depend on the coordinates $\rho$ and $z$. Each of the functions $U_i$, $i=1,2,3$ solves the three-dimensional Laplace equation
\be\n{1.2}
U_{i,\rho\rho}+\frac{1}{\rho}U_{i,\rho}+U_{i,zz}=0\,~,
\ee
such that the following constraint holds:
\be\n{1.3}
U_1+U_2+U_3=\ln \rho\,.
\ee
Here, and in what follows $(...)_{,a}$ stands for the partial derivative of the expression $(...)$ with respect to the coordinate $x^a$.

If the functions $U_i$, $i=1,2,3$ are known, the function $\nu$ can be derived by simple line integral from the following equations:
\ba
\nu_{,\rho}&=&-\rho(U_{1,\rho}U_{2,\rho}+U_{1,\rho}U_{3,\rho}+U_{2,\rho}U_{3,\rho}\non\\
&-&U_{1,z}U_{2,z}-U_{1,z}U_{3,z}-U_{2,z}U_{3,z})\,,\n{1.4a}\\
\nu_{,z}&=&-\rho(U_{1,\rho}U_{2,z}+U_{1,\rho}U_{3,z}+U_{2,\rho}U_{3,z}\non\\
&+&U_{1,z}U_{2,\rho}+U_{1,z}U_{3,\rho}+U_{2,z}U_{3,\rho})\,.\n{1.4b}
\ea

Here, we shall consider a five-dimensional Weyl solution representing a background Weyl solution defined by $\tu_i$, $i=1,2,3$, and $\tnu$, which is distorted by external, static, axisymmetric fields defined by $\hu_i$, $i=1,2,3$, and $\hnu$. The metric functions of the corresponding spacetime can be defined as follows: 
\be\n{1.5}
U_i:=\tu_i+\hu_i\hhh \nu:=\tnu+\hnu\,,
\ee
where according to the constraint (\ref{1.3}) we have
\be\n{1.6}
\tu_1+\tu_2+\tu_3=\ln \rho\hhh \hu_1+\hu_2+\hu_3=0\,.
\ee
In what follows, we consider distortion by external gravitational fields due to remote masses that have an axisymmetric configuration with respect to the axes corresponding to the Killing vectors  
$\xi^\alpha_{(\chi)}=\delta^\alpha_{\,\,\, \chi}$ and $\xi^\alpha_{(\phi)}=\delta^\alpha_{\,\,\, \phi}$. To present the ansatz, accordingly, we define
\ba
&&\hspace{-0.7cm}\tu_1:=\tu+\tw+\ln\rho\hhh\tu_2:=-\tw\hhh\tu_3:=-\tu\,,\n{1.7a}\\
&&\hspace{1.5cm}\tnu:=\tv+\tu+\tw\,,\n{1.7b}\\
&&\hspace{-0.3cm}\hu_1:=\hu+\hw\hhh\hu_2:=-\hw\hhh\hu_3:=-\hu\,,\n{1.7c}\\
&&\hspace{1.5cm}\hnu:=\hv+\hu+\hw\,.\n{1.7d}
\ea
Then, the metric (\ref{1.1}) takes the following generalized Weyl form \cite{ERW}:
\ba\n{1.8}
ds^2&=&e^{2(\tu+\tw+\hu+\hw)}[-\rho^2dt^2+e^{2(\tv+\hv)}(dz^2+d\rho^2)]\non\\
&+&e^{-2(\tw+\hw)}d\chi^2+e^{-2(\tu+\hu)}d\phi^2\,.
\ea
The background fields $\tu$ and $\tw$ satisfy the three-dimensional Laplace equation (\ref{1.2}), and the function $\tv$ can be obtained via a simple line integral from the following equations:
\ba
\tv_{,\rho}&=&\rho\,(\tu_{,\rho}^2+\tw_{,\rho}^2+\tu_{,\rho}\tw_{,\rho}-\tu_{,z}^2-\tw_{,z}^2-\tu_{,z}\tw_{,z})\,,\non\\
\n{1.9a}\\
\tv_{,z}&=&\rho\,(2\tu_{,\rho}\tu_{,z}+2\tw_{,\rho}\tw_{,z}+\tu_{,\rho}\tw_{,z}+\tu_{,z}\tw_{,\rho})\,.\n{1.9b}
\ea
The distortion fields $\hu$ and $\hw$ satisfy the three-dimensional Laplace equation (\ref{1.2}), and the function $\hv$ representing interaction between the distortion fields can be derived by the line integral from the following equations:
\ba
\hv_{,\rho}&=&\rho\,(\hu_{,\rho}^2+\hw_{,\rho}^2+\hu_{,\rho}\hw_{,\rho}-\hu_{,z}^2-\hw_{,z}^2-\hu_{,z}\hw_{,z}\non\\
&+&\tu_{,\rho}\hw_{,\rho}+\tw_{,\rho}\hu_{,\rho}-\tu_{,z}\hw_{,z}-\tw_{,z}\hu_{,z}\non\\
&+&2[\tu_{,\rho}\hu_{,\rho}+\tw_{,\rho}\hw_{,\rho}-\tu_{,z}\hu_{,z}-\tw_{,z}\hw_{,z}])\,,
\n{1.10a}\\
\hv_{,z}&=&\rho\,(2\hu_{,\rho}\hu_{,z}+2\hw_{,\rho}\hw_{,z}+\hu_{,\rho}\hw_{,z}+\hu_{,z}\hw_{,\rho})\non\\
&+&\tu_{,\rho}\hw_{,z}+\tu_{,z}\hw_{,\rho}+\tw_{,\rho}\hu_{,z}+\tw_{,z}\hu_{,\rho}\non\\
&+&2[\tu_{,\rho}\hu_{,z}+\tu_{,z}\hu_{,\rho}+\tw_{,\rho}\hw_{,z}+\tw_{,z}\hw_{,\rho}])
\,.\n{1.10b}
\ea

In the following section, we use the generalized Weyl form (Emparan-Real metric) \cite{ERW}, and this ansatz \cite{13s} to construct a metric representing a five-dimensional distorted black ring and study its properties. 
 
\section{Black Ring}

In \cite{13s}, an ansatz was presented for deriving a distorted five-dimensional black object with three commuting, orthogonal Killing vector fields based on the generalized Weyl form \cite{ERW}. Here, we review this ansatz. The Weyl form of the metric is given by the functions $U_{1..3}$ and $\nu$ in $\rho$ and $z$ coordinates \cite{ERW}:
\ba
&&e^{2U_1}=\frac{R_3-\xi_3}{R_2-\xi_2}~,\\
&&e^{2U_2}=\frac{R_1-\xi_1}{A}~,\\
&&e^{2U_3}=\frac{(R_1+\xi_1)(R_2-\xi_2)}{A(R_3-\xi_3)}~,\\
&&e^{2\nu}=\frac{1+\mu}{4A}\frac{Y_{23}}{R_1 R_2 R_3}\sqrt{\frac{Y_{12}}{Y_{13}}}\sqrt{\frac{R_2-\xi_2}{R_3-\xi_3}}
\ea
where 
\ba
&&\xi\equiv z-c_i~,\\
&&R_1\equiv \sqrt{\rho^2+\xi_1^2}~,\\
&&R_2\equiv \sqrt{\rho^2+\xi_2^2}~,\\
&&R_3\equiv -\sqrt{\rho^2+\xi_3^2}~,\\
&&Y_{ij}\equiv R_i R_j +\xi_i \xi_j +\rho^2~,
\ea
from which it follows that $U_1$ is the Newtonian potential produced by a finite rod $-\mu/(2A)\leq
z\leq\mu/(2A)$, $U_2$ is the potential produced by a semi-infinite rod $z\geq 1/(2A)$, and $U_3$ is the
potential produced by a semi-infinite rod $z\leq-\mu/(2A)$ and a finite rod $\mu/(2A)\leq z \leq1/(2A)$.
For $\mu=1$, these sources reduce to those of the five-dimensional Schwarzschild solution. 
Here we have, with $\alpha=A$, to reproduce the Weyl form of the metric:
\ba\label{c}
c_1=\alpha/(2A^2)=1/(2A),~~~~c_2=\alpha\mu/(2A^2)=\mu/(2A),~~~~c_3=-\alpha\mu/(2A^2)=-\mu/(2A)~.
\ea
The parameters $\mu$ and $A$ will be taken to lie in the range $0\leq\mu\leq1$, $A>0$. 
This metric can also be presented in a more simple form:  
\be
ds^2=-\frac{F(x)}{F(y)}dt^2+\frac{1}{A^2(x-y)^2}\biggl[\biggl(\frac{F(y)^2}{1-x^2} dx^2+\frac{F(x) F(y)}{y^2-1}dy^2\biggl)+F(x)(y^2-1)d\psi^2+\frac{F(y)^2}{F(x)}(1-x^2) d\phi^2\biggl],\n{METRIC1}
\ee
where $F(x)=1-\mu x$ and $F(y)=1-\mu y$. The coordinate $x$ is in
the range $-1\leq x \leq 1$, and the coordinate $y$ is in the range $y\leq -1$ with the black hole horizon located at $y\rightarrow -\infty$. The relation between the coordinates $(x,y)$ and $(\rho,z)$ is given by the following transformations:
\ba
&&\rho=\frac{\alpha}{A^2(x-y)^2}\sqrt{F(x)F(y)(1-x^2)(y^2-1)}~,\n{Trans1}\\
&&z=\frac{\alpha(1-xy)(F(x)+F(y))}{2A^2(x-y)^2}~.\n{Trans2}
\ea
The functions $\tu$, $\tw$, and $\tv$ in the coordinates $(\rho,z)$ are the following:
\ba
&&e^{2\tu(\rho,z)}=\frac{A(R_3-\xi_3)}{(R_2-\xi_2)(R_1+\xi_1)}~,\\
&&e^{2\tw(\rho,z)}=\frac{A}{R_1-\xi_1}~,\\
&&e^{2\tilde{V}(\rho,z)}=\frac{1+\mu}{4A^3}\frac{Y_{23}}{R_1 R_2 R_3}\sqrt{\frac{Y_{12}}{Y_{13}}}(R_1^2-\xi_1^2)({\frac{R_2-\xi_2}{R_3-\xi_3}})^{3/2}~. 
\ea
Note, that $y=-1$ is seen as the origin of polar coordinates, and hence $y$ cannot be continued beyond $-1$. Returning to the metric \eqref{METRIC1}, it is now clear that the topology of the horizon is $S^1\times S^2$, which
justifies calling this solution a black ring. $x$ is the polar
coordinate on the $S^2$; it follows that $x=-1$ points away from the ring and $x=1$ points into the hole in the center of the ring.  It is clear from the metric  \eqref{METRIC1} that the only values of $x$ and $y$ that can correspond to asymptotic
infinity are $x=y=-1$. The black ring has a conical singularity either at $x=1$ or at $x=-1$. If the asymptotic metric does not contain a conical singularity, then the mass of the black ring is  
\be
M =\frac{3\pi\mu(1+\mu)}{4G_5 A^2}\,,
\ee
where $G_5$ is  Newton's constant in five dimensions. For  fixed  $A$, a change in $\mu$ changes   the black ring's mass. On the other hand, if the deficit membrane extends to infinity, the mass of the ring is 
\be
M=\frac{3\pi\mu\sqrt{1-\mu^2}}{4G_5 A^2}\,.
\ee
\begin{center}
\begin{figure}[t]
\centering
  \includegraphics[width=8cm]{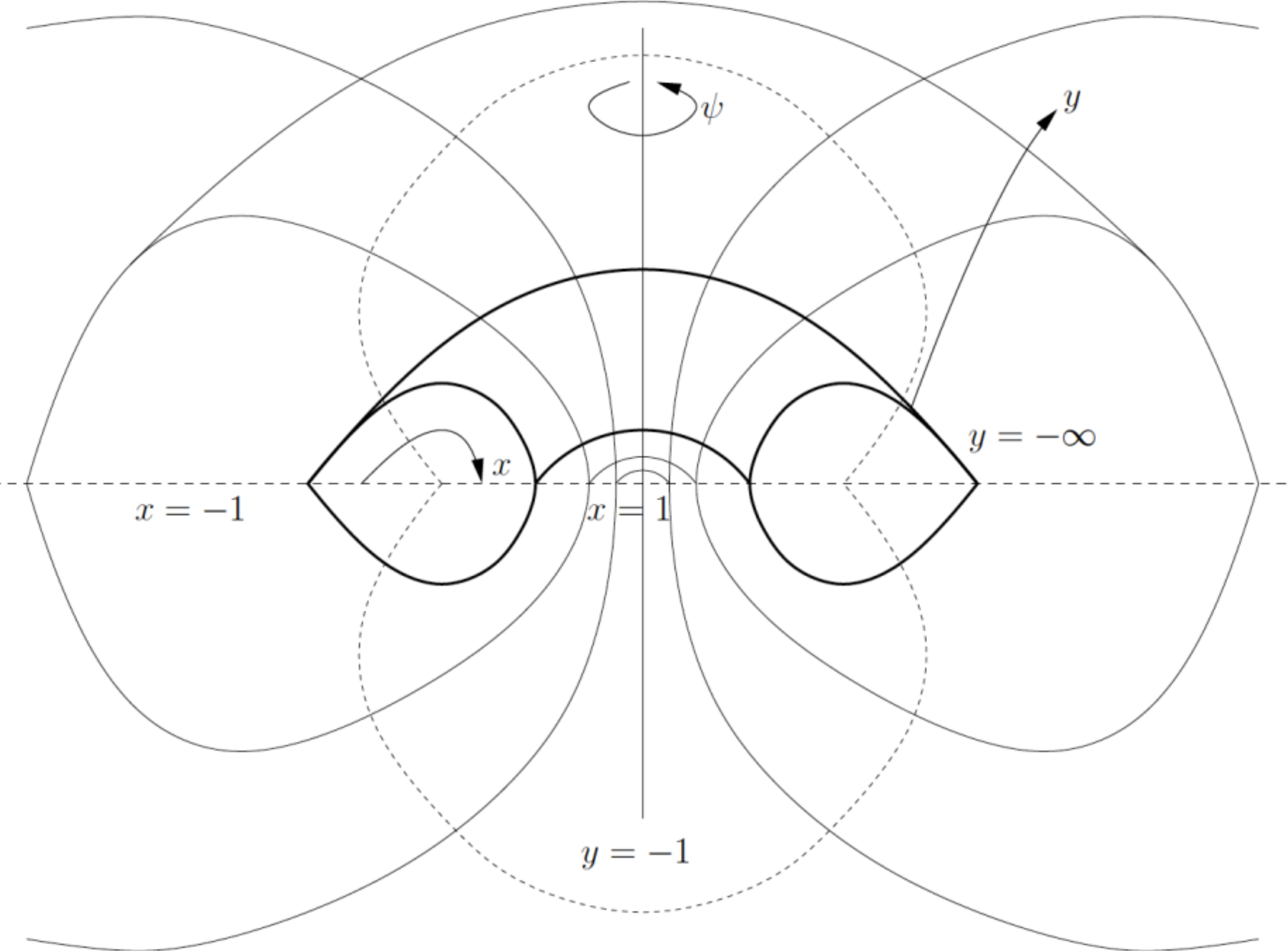}~~~~~~~
   \caption{Spatial sections of the black ring metric. The coordinate $\phi$ is suppressed. The surfaces of constant $x$ are denoted by dotted lines. The surfaces of constant $y$ are nested surfaces of topology $S^2\times S^1$. The horizon is at $y=-\infty$. The surface at $y=-1$ degenerates into an axis of rotation where the orbits of $\psi$ shrink to zero. The local distorted black ring solution is defined in the region $y\ll -1$, near the horizon and far away from the axis $y=-1$. Image credit:\cite{ERW} }
\end{figure}
\end{center}

\section{Metric of a distorted black ring}

In the previous section, we reviewed \cite{ERW}, demonstrating that the five-dimensional black ring solution can be written in the generalized Weyl form. Here, we present a metric describing a five-dimensional static vacuum black ring distorted by external fields. The fields' sources are located at asymptotic infinity and are not included in the metric. As a result, the corresponding spacetime is not asymptotically flat. We consider the spacetime near the regular black hole horizon, far away from the sources. In this case, the solution represents a {\em local black hole} in analogy with a four-dimensional distorted vacuum black hole studied in \cite{dis3}. We focus on the study of the spacetime near the black hole horizon. The corresponding metric ansatz of such a black ring is given by
\ba
ds^2&=&-e^{2(\hu+\hw)}\frac{F(x)}{F(y)}dt^2+\frac{1}{A^2(x-y)^2}\biggl[e^{2(\hat{V}+\hu+\hw)}\biggl(\frac{F(y)^2}{1-x^2} dx^2+\frac{F(x) F(y)}{y^2-1}dy^2\biggl)\nonumber\\
&+&e^{-2\hw}F(x)(y^2-1)d\psi^2+e^{-2\hu}\frac{F(y)^2}{F(x)} (1-x^2)d\phi^2\biggl]~.\n{MetricA}
\ea
To derive the solution it is easier to start from the $(\rho,z)$ coordinates. The next step is to find the functions $\hu$, $\hw$, and $\hat{V}$. Once these functions are known in $(\rho,z)$, based on the transformation (\ref{Trans1}) and (\ref{Trans2}) the functions $\hu$, $\hw$, and $\hat{V}$ in $x$ and $y$ coordinates can be derived. We do not write the functions explicitly in the $x$ and $y$ coordinates, due to their complexity. 

We start with the Laplace equation (\ref{1.2}) for the distortion fields $\hu$ and $\hw$. In the cylindrical coordinates $(\rho,z)$ the solution is well known and has the following form:
\ba\n{3.1}
\hx(\rho,z)=\sum_{n\geq0}\left[A_n\,R^{n}+B_n\,R^{-(n+1)}\right]P_n(\cos\vartheta)\,,
\ea
where 
\ba\n{3.2}
R=\frac{\sqrt{\rho^2+z^2}}{m}\hhh\cos\vartheta=z/R\,,
\ea
and $P_n(\cos\vartheta)$ are the Legendre polynomials of the first kind. In what follows, $\hx$ is either $\hu$ or $\hw$. The coefficients $A_n$ and $B_n$ in the expansion \eq{3.1} correspond to interior and exterior multipole moments, respectively \cite{Multipoles1,Multipoles2}. We consider only the distortion fields defined by $A_n$'s which describe a local black ring distorted by external sources. For $\hu$, $A_n$'s are named $a_n$ and for $\hw$, $A_n$'s are named $b_n$.  We shall simply call the $a_n$ and $b_n$ coefficients {\em multipole moments}. 

We note that our solution is valid for $R <  R^\prime$, where $R'$ is the characteristic length scale at which
the external sources are located (i.e., outside the the region where our  solution is valid).  
We consider  distortions with $|a_n|<1$ and $|b_n|<1$. For large values of $a_n$ and $b_n$, the gravitational effect of the distorting fields dominates over that of the black ring. Physically, a small multiple moment corresponds to distortion of the black object distorted by masses smaller than the object itself, whereas a large multiple moment would result from surrounding the black object with masses much larger than the object itself.  Another way to see this is that for large multipole moments the quantity $\exp{2U}$ is so large that it dominates the spacetime gravitational potential.   
 Referring to Fig. 1, the local solution can be considered valid in the interior region of one of the surfaces of constant $y$.

If the distortion fields $\hu$ and $\hw$ are known, the function $\hv$ can be derived from Eqs. (\ref{1.10a}) and (\ref{1.10b}). For the function $\hv$, we write 
\ba
&&\hat{V}=\hat{V}_1+\hat{V}_2,\n{formV1}\\
&&\hat{V}_1=\hat{V}_{\hu\hu}+\hat{V}_{\hu\hw}+\hat{V}_{\hw\hw},\n{formV2}\\
&&\hat{V}_2=2\hat{V}_{\tu \hu}+\hat{V}_{\tu \hw}+\hat{V}_{\tw \hu}+2\hat{V}_{\tw \hw},\n{formV3}
\ea
where each of the $\hat{V}_{(fg)}$'s is given by the solution of 
\ba
&&\hat{V}_{(fg),\rho}=\rho(f_{,\rho}g_{,\rho}-f_{,z}g_{,z})~,\n{Vdp1}\\
&&\hat{V}_{(fg),z}=\rho(f_{,\rho}g_{,z}+f_{,z}g_{,\rho})~.\n{Vdp2}
\ea
Note that $(fg)$ is the index, while comma $\rho$ and $z$ are the partial derivatives with respect to $\rho$ and $z$. This form of function $\hat{V}$ [Eqs. (\ref{formV1})-(\ref{formV3})] is read from Eqs. (\ref{1.10a}) and (\ref{1.10b}). The three parts $\hat{V}_{\hu\hu}$, $\hat{V}_{\hw\hu}$, and $\hat{V}_{\hw\hw}$ involve only the distortion fields. We can write the function $\hat{V}_1$ in the following form:
\ba
\hat{V}_1=\hat{V}_{\hu\hu}+\hat{V}_{\hu\hw}+\hat{V}_{\hw\hw}=\sum_{n,k\geq1}\frac{nk}{n+k}(a_na_k+a_nb_k+b_nb_k)R^{n+k}
[P_nP_k-P_{n-1}P_{k-1}]\,,\n{3.4c}
\ea
where $P_n\equiv P_n(z/R)$. The four parts $\hat{V}_{\tu \hu}$, $\hat{V}_{\tu \hw}$, $\hat{V}_{\tw \hu}$, and $\hat{V}_{\tw \hw}$ involve the interaction of the distortion fields with the background fields $\tu$ and $\tw$. Note that the distortion fields $\hu$ and $\hw$ are independent in general. Therefore, one can consider $\hu=0$ or $\hw=0$, in which case the corresponding part in $\hat{V}$ has to be taken to zero. However, further conditions on the metric may impose a relation between the distortion fields. Due to the fact that $\tu$ and $\tw$ are logarithmic functions, we can further decompose Eqs. (\ref{1.10a}) and (\ref{1.10b}) for the derivation of $\hv_2$ in the following manner:
\ba
&&\hat{V}_ {\tw \hx}=-\hat{V}_{\hat X R_{1-}}~,\\
&&\hat{V}_{\tu\hx}=-\hat{V}_{\hat X R_{1+}}-\hat{V}_{\hat X R_{2-}}+\hat{V}_{\hat X R_{3}}, 
\ea where corresponding to each term, Eqs. (\ref{1.10a}) and (\ref{1.10b}) are decomposed in the forms 
(\ref{Vdp1}) and (\ref{Vdp2}) with the following notation: For $\hat{V}_{\hat X R_{1-}}$, $f=\hx$ and $g=1/2\ln{[(R_1-\xi_1)/A]}$. For $\hat{V}_{\hat X R_{1+}}$, $f=\hx$ and $g=1/2\ln{[(R_1+\xi_1)/A]}$. For $\hat{V}_{\hat X R_{2-}}$, $f=\hx$ and $g=1/2\ln{[(R_2-\xi_2)/A]}$. For $\hat{V}_{\hat X R_{3}}$, $f=\hx$ and $g=1/2\ln{[(R_3-\xi_3)/A]}$. Then, each term can be found by a line integral
\be\n{2.17}
\hat{V}_{\square\square}(\rho,z)=\int_{(\rho_0,z_0)}^{(\rho,z)}\left[\hat{V}_{\square\square,z'}(\rho',z')dz'+\hat{V}_{\square\square,\rho'}(\rho',z')d\rho'\right]\,,
\ee
where the integral is taken along any path connecting the points $(\rho_0,z_0)$ and $(\rho,z)$. Thus, the field $\hv$ is defined up to an arbitrary constant of integration defined by the point $(\rho_0,z_0)$. We choose this arbitrary constant using a boundary condition. Here, $\square$'s are to be filled with the corresponding notation for each term in $\hv_2$. 

Thus, we have 
\ba
&&\hat{V}_{\hx R_{i\pm}}=\int\frac{\rho}{2R_i}\biggl( \frac{\rho}{{R_i\pm\xi_i}}\hx_{,\rho} \mp \hx_{,z}\biggl) d\rho+\sum_{n \geq 1} d_n \frac{z^n}{2m^n}+C\,~~~~~i=1,2 \\
&&\hat{V}_{\hx R_{3}}=\int\frac{\rho}{2R_3}\biggl( \frac{\rho}{{R_3-\xi_3}}\hx_{,\rho} + \hx_{,z}\biggl) d\rho+\sum_{n \geq 1} d_n \frac{z^n}{2m^n}+C\,, 
\ea
where $i=1$ or $2$, and $d_n$ is $a_n$ or $b_n$. Therefore, we can derive $\hv_2$ as the following for each of the multiple moments $n=1..4$:
\ba
\hv_2&=&\frac{1}{2z}\biggl[a_1{(R_1-2R_2+2R_3)} -b_1{(R_1+R_2-R_3)}-3(a_1+b_1)z \biggl]R P_1,~~~~~n=1,\\
\hv_2&=&-\frac{3}{2}(a_2+b_2)R^2P_2+\frac{1}{2m}\biggl[\biggl(a_2(R_1-2R_2+2R_3)-b_2(R_1+R_2-R_3)\biggl)\nonumber\\
&+&\frac{1}{z}\biggl(a_2{(R_1c_1-2R_2c_2+2R_3c_3)}-b_2{(R_1c_1+R_2c_2-R_3c_3)}\biggl)\biggl]R P_1,~~~~~n=2\\
\hv_2&=&-\frac{3}{2}(a_3+b_3)R^3P_3+\frac{1}{4z^3}\biggl(-a_3(R_1^3-2R_2^3+2R_3^3)+b_3(R_1^3+R_2^3-R_3^3)\biggl)R^3P_1^3\nonumber\\
&+&\frac{3}{4zm^2}\biggl[a_3\biggl ( (R_1-2R_2+2R_3)z^2+R_1c_1^2-2R_2 c_2^2+2R_3 c_3^2\biggl)\nonumber\\
&+&b_3\biggl(-(R_1+R_2-R_3)z^2-R_1c_1^2-R_2 c_2^2+R_3 c_3^2\biggl)\biggl]RP_1,~~~~~~n=3
\ea 
\ba
\hv_2&=& -\frac{3}{2} (a_4+b_4) R^4 P_4 -\frac{1}{z^3}\biggl( a_4 (R_1^3-2R_2^3+2R_3^3)-b_4(R_1^3+R_2^3-R_3^3)\biggl)R^4P_1^4 \nonumber\\
&+&\frac{1}{m^3}\biggl[\left(3a_4 (\frac{1}{2} R_1-R_2+R_3)-\frac{3b_4}{2}(R_1+R_2-R_3)\right)z^2\nonumber\\
&+&3\left(-a_4(\frac{1}{2}R_1 c_1-R_2 c_2+R_3 c_3) +\frac{b_4}{2}(R_1c_1+R_2 c_2-R_3 c_3)\right)z\nonumber\\
&+& \frac{a_4 r^2}{4 z}\left((R_1-2R_2+2R_3)z-R_1 c_1+2R_2 c_2-2R_3 c_3\right)\nonumber\\
&-& \frac{b_4 r^2}{4 z}\left((R_1+R_2-R_3)z-R_1 c_1-R_2 c_2+R_3 c_3\right)\nonumber\\
&+&\frac{1}{2 z} \left(a_4(R_1 c_1^3-2R_2 c_2^3+2 R_3 c_3^3)-b_4(R_1 c_1^3+R_2 c_2^3-R_3 c_3^3)\right)\nonumber\\
&+&\frac{3}{2}\left(a_4(R_1c_1^2-2R_2c_2^2+2R_3c_3^2)-b_4(R_1c_1^2+R_2 c_2^2-R_3 c_3^2)\right)\biggl] R P_1\,~~~~~~n=4\,. 
\ea 
Note that in practice, for the analysis of the distorted black ring we restrict ourselves only up to $n=2$.
\section{General Conditions}
The metric of a constant $t$ slice through the horizon (i.e., horizon surface) is
\be
ds^2=\frac{1}{A^2}\biggl(F(x)e^{-2\hw_0}d\psi^2+\mu^2e^{2(\hv_0+\hu_0+\hw_0)}\frac{dx^2}{1-x^2}+\mu^2e^{-2\hu_0}\frac{1-x^2}{F(x)}d\phi^2\biggl). 
\ee
Here, $\hu_0$, $\hw_0$, and $\hv_0$ are calculated on the horizon. 
One can see that the distortion fields $\hu, \hw$, and $\hv$
are smooth on the black ring horizon. Thus, the horizon is regular, and
this solution represents a local black ring distorted by
the external static fields.  

We shall regard the distorted black ring Eq. (\ref{MetricA}) as a local solution, valid only in a certain neighborhood of the horizon. The external sources distorting the black hole are located beyond this neighborhood where the spacetime is not vacuum and the solution is not valid. In other words, even though our metric represents a vacuum spacetime, some matter sources exist exterior to the region of validity of the solution and cause distortion of the black hole. A global solution can be constructed by extending the metric to an asymptotically flat solution by a sewing technique. This can be realized by cutting the spacetime manifold in the region where the metric is valid and attaching to it another spacetime manifold where the solution is not vacuum anymore, but the sources of the distorting matter are also included. Beyond this nonvacuum region, we assume to have an asymptotically flat vacuum solution. 

For the metric (\ref{MetricA}), the $\{tt\}$ component of the Einstein equations reads
\ba\n{3.7}
R_{\alpha\beta}\delta^\alpha_{\,\,\, t}\delta^\beta_{\,\,\, t}&=&\rho^2e^{-2\hv}\left(\triangle\hu+\triangle\hw\right)\non\\
&=&8\pi\left(T_{\alpha\beta}-\frac{T^\gamma_{\,\,\,\,\gamma}}{3}g_{\alpha\beta}\right)\delta^\alpha_{\,\,\, t}\delta^\beta_{\,\,\, t}\,,
\ea
where $T_{\alpha\beta}$ is the energy-momentum tensor representing the sources, and $\triangle$ is the Laplace operator. Strong energy conditions impose a condition on the fields $\hu$ and $\hw$. If the sources of the distortion were to be included in the solution, the Einstein equations would not be vacuum. This would require continuing the solution beyond the internal vacuum region. Note that for the following analysis, it is not necessary to formally implement this procedure. If the sources satisfy the strong energy condition, the right-hand side of Eq. (\ref{3.7}) must be non-negative. The Laplace operator $\triangle$ is a negative operator; hence, the strong energy condition implies
that
\ba
\hu+\hw\leq 0
\ea
assuming that $\hu+\hw=0$ at asymptotically flat infinity. 

One can use the analogy with the Newtonian picture in order to give an interpretation for multiple moments, which clarifies that higher-order multiple moments are expected to be smaller than the lower-order multiple moments. In Sec. VI, we restrict our analysis to $n=0...2$. The case with $a_1\neq 0$ and $a_2=0$ is called the dipole distortion. The case with with $a_1= 0$ and $a_2\neq 0$ is called the quadrupole distortion. The case with $a_1\neq 0$ and $a_2\neq 0$ is called the dipole-quadrupole distortion.

Boundary values of the distortion functions $\hu,~\hw$ on the horizon are given by
\ba
&&\hu_0=\hu(x)|_H=\sum_{n\geq0}a_{n}x^n,\\ \n{huHorizon}
&&\hw_0=\hw(x)|_H=\sum_{n\geq0}b_{n}x^n,
\ea
where we have chosen $m=\mu/2A$. Note that $m$ is an arbitrary constant. This choice corresponds to a rescaling of the multiple moments; we shall not rename
them. The boundary value of the distortion function $\hv$ on the horizon is
\ba
\hv_0=\hv(x)|_H=-3\sum_{n\geq 1}^4(a_{n}+b_{n})x^{n}+\sum_{n\geq 1}^2(2a_{2n}+b_{2n})-\sum_{n\geq 1}^4\frac{(a_{n}-b_{n})}{2\mu^n}+C\, . \n{hvHorizon}
\ea
and in what follows, we shall set $C=0$ without loss of generality; solutions with different values of constant $C$ can
be related to this solution by rescaling the parameter $A$,  and the periods of $\psi$ and $\phi$.

Another condition that we wish to formulate is the conical regularity condition. The metric has no conical singularities  along the semiaxes $\theta=0$ and $\theta=\pi$ which correspond to angular deficit or excess, if the space there is locally flat. Consider the $x\phi$ part of the metric, which is conformal to
\be
ds_{x\phi}^2=e^{2(\hv+2\hu+\hw)}\frac{dx^2}{1-x^2}+\frac{1-x^2}{F(x)}d\phi^2\,. 
\ee
Let $x=-\cos\theta$ with $0\leq \theta \leq \pi$. This gives 
\be
ds_{x\phi}^2=e^{2(\hv+2\hu+\hw)}d\theta^2+\frac{\sin^2\theta}{1+\mu\cos\theta}d\phi^2\,. 
\ee
If the period of $\phi$ is considered to be $\Delta_\phi$, we will have conical regularity (no conical singularity) on the semiaxis $\theta=0$, provided 
\ba
{2\pi}{\sqrt{1+\mu}}\,{e^{\hv+2\hu+\hw}\rvert_{\theta=0}}={\Delta_\phi}\,,\n{2.21a}
\ea 
whereas the regularity condition for the semiaxis $\theta=\pi$ reads
\be
{2\pi}{\sqrt{1-\mu}}\,{e^{\hv+2\hu+\hw}\rvert_{\theta=\pi}}={\Delta_\phi}\,.\n{2.21b}
\ee
The conditions (\ref{2.21a}) and (\ref{2.21b}) guarantee that the space is locally 
regular along the axis for all $y$, by requiring that the ratio of
the circumference to the radius of an infinitesimal circle, drawn orthogonally to the axis, be $2\pi$ when the period of $\phi$
is chosen as $\Delta_\phi$. The conical regularity conditions (\ref{2.21a}) and (\ref{2.21b}) combined read
\be
\frac{e^{\hv+2\hu+\hw}\rvert_{\theta=\pi}}{e^{\hv+2\hu+\hw}\rvert_{\theta=0}}=\frac{\sqrt{1+\mu}}{\sqrt{1-\mu}}\,.\n{Noconical0}
\ee
In the case of an undistorted black ring, there exists a conical singularity in one of the poles. In the full metric, this singularity is extended in two other spatial directions and describes a ``deficit membrane," which is the five-dimensional analogue of a four-dimensional deficit string \cite{ERW}. For the general distortions $\hv$, $\hu$ and $\hw$, $(\hv+2\hu+\hw)\rvert_{\theta=\pi}$ depends on the values of $y$, which makes it impossible to remove the conical singularity by adjusting the multiple moments. However, for $\hw=-\hu/2$, the left-hand side of Eq. (\ref{Noconical0}) is independent of $y$, and we can adjust the multipole moments to remove the conical singularity that exists in an undistorted black ring solution. Therefore, in what follows, we consider $\hw=-\hu/2$. For the distortion including $n=0...4$, Eq. (\ref{Noconical0}) reads
\ba\n{Noconical}
\frac{e^{\hv+2\hu+\hw}\rvert_{\theta=\pi}}{e^{\hv+2\hu+\hw}\rvert_{\theta=0}}=\exp\biggl[-\frac{3}{2}(1-\mu)\sum_{n=1}^4 \sum_{k=n}^4 \frac{a_k}{\mu^n}\biggl]=\frac{\sqrt{1+\mu}}{\sqrt{1-\mu}}.
\ea For the dipole distortion ($a_{n>1}=0$) in Eq. (\ref{Noconical}), we get the following relation between $a_1$ and the parameter $\mu$ ,
\be\n{Noconicald}
a_1=\frac{\mu}{3(1-\mu)} \ln\left(\frac{1-\mu}{1+\mu}\right).
\ee For the quadrupole distortion ($a_1=0,a_2\neq 0$, $a_{n>2}=0$) in Eq. (\ref{Noconical}), we get the following relation relation between $a_2$ and the parameter $\mu$:
\be\n{Noconicalq}
a_2=\frac{\mu^2}{3(1-\mu^2)} \ln\left(\frac{1-\mu}{1+\mu}\right) \,. 
\ee
Equations (\ref{Noconicald}) and (\ref{Noconicalq}) satisfy Eq. (\ref{Noconical}), which makes sure we have Eqs. (\ref{2.21a}) and (\ref{2.21b}) satisfied. The undistorted black ring has conical singularities describing a deficit membrane that either extends to infinity or forms a disc inside the ring, which prevents the ring from collapsing to form a spherical black hole horizon. In the latter case, the solution is asymptotically flat. Here, we see that a static ring could be free of the conical singularities due to the presence of the external matter sources, which prevent the ring from collapsing to form a spherical black hole horizon. Tuning at least one of the multipole moments, we can remove the conical singularity that exists in the undistorted black ring solution. 

Let us consider the $(y,\psi)$ section of the metric (\ref{MetricA}). As $y\rightarrow -1$, $g_{\psi\psi}$ tends to zero. To analyze this, we set $y=-\cosh(\xi/\sqrt{1+\mu})$. Near $\xi=0$, the
$\xi\psi$ part of the metric is conformal to 
\be
ds_{\xi\psi}^2=e^{-2(\hv+\hu+2\hu)|_{\xi=0}}d\xi^2+\frac{\xi^2}{1+\mu} d\psi^2, 
\ee
The conical regularity condition would require that 
\be
e^{-2(\hv+\hu+2\hu)|_{\xi=0}} \frac{\Delta_\psi}{\sqrt{1+\mu}}=2\pi.\n{ConicalPsi}
\ee
It is not possible to satisfy Eq. (\ref{ConicalPsi}) in general, since the term $(\hv+\hu+2\hu)|_{\xi=0}$ depends on $x$.  However, the solution can be considered valid in the interior region of one of the surfaces of $y=const.$, between the horizon at $y$ equal to $-\infty$ and $y\ll -1$. The matter sources are located in an intermediate region between this $y=const.$ surface and $y=-1$. 
We can therefore identify $\psi$ with period $\Delta_\psi=2\pi$, without creating any conical singularity
in our local black hole solution, which is defined in the area near the horizon (recall that infinity is at $x=y=-1$).

\section{Analysis}

\begin{center} 
\begin{figure}[t]
\centering
  \includegraphics[width=7cm]{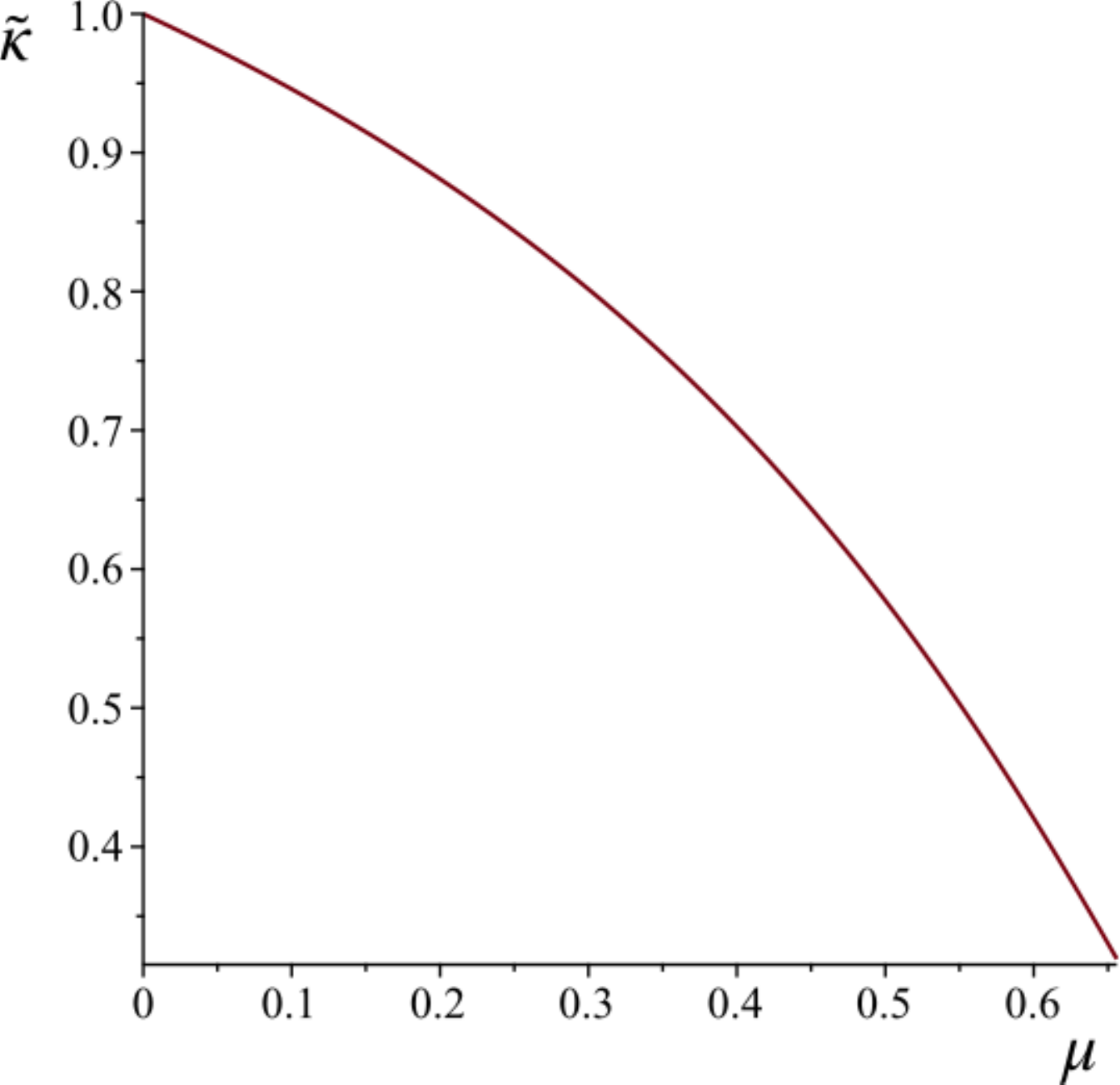}~~~~~~~
   \includegraphics[width=7cm]{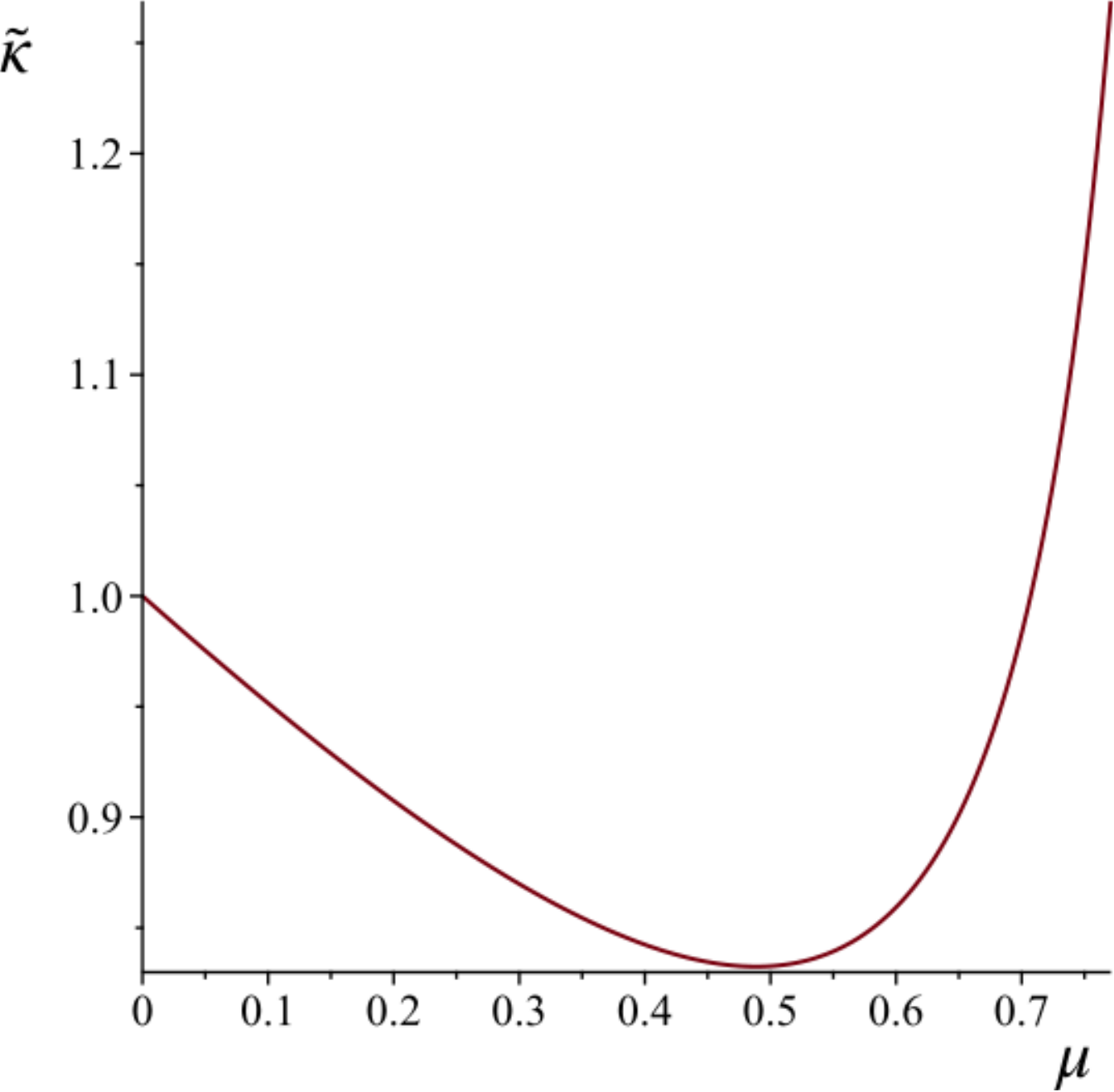}
   \caption{Behavior of $\tilde{\kappa}$ with respect to parameter $\mu$. On the left, we have plotted the behavior of $\tilde{\kappa}$ for a dipole multiple moment with $\mu \in (0..0.656)$. On the right, we have plotted the behavior of $\tilde{\kappa}$ for a quadrupole multiple moment with $\mu \in (0..0.771)$. The upper limits for the parameter $\mu$ are chosen such that the resulting multiple moments (dipole or quadrupole) are less than $-1$.\label{kappa}}
\end{figure}
\end{center}

The distorted black hole horizon is defined by $y\rightarrow -\infty$. The area of the horizon is 
\be
\mathcal{A}_H=\frac{2\mu^2}{A^3}\Delta_\psi\Delta_\phi=\frac{8\pi^2\mu^2}{A^3}\sqrt{1+\mu}~e^{\hv+2\hu+\hw|_{\theta=0}}~\,.
\ee The area of the horizon for the first four orders is the following:
\be
\mathcal{A}_H=\frac{8\pi^2\mu^2}{A^3}\sqrt{1+\mu}~\exp\left(\frac{3}{2}(a_0-a_1-a_3)+\frac{3}{4}\sum_{n=1}^{4}\frac{a_n}{\mu^n}\right)\,. 
\ee
The surface gravity of a distorted black ring is given by
\be
\kappa=\frac{A}{2\mu}\exp\biggl[\sum_{n=1}^{4}{2^{n-1}\left(\frac{A^2}{\alpha\mu}\right)^n\left[a_n\left(c_1^n-2c_2^n-2c_3^n\right)-b_n\left(c_1^n+c_2^n+c_3^n\right)\right]}\biggl] \, . 
\ee If we take into account that $\alpha=A$, Eq. (\ref{c}), and the conical regularity condition ($b_n=-a_n/2$), we get, 
\be
\kappa=\frac{A}{2\mu}\exp\biggl[\sum_{n=1}^{4}{\frac{3}{4}a_n\left(\mu^{-n}-(-1)^n-1\right)}\biggl] \, .
\ee If we consider dipole-quadrupole, we get the following relationship for $\tilde{\kappa}=\kappa/\kappa_{iso}$:

\be
\tilde{\kappa} = e^{\frac{-3(2\mu^2a_2-\mu a_1-a_2)}{4\mu^2}}
\ee
Here, we should note that $\kappa_{iso}={A}/(2\mu)$ is the surface gravity of the undistorted black ring.
In Fig. \ref{kappa}, we have plotted the behavior of $\tilde{\kappa}=\kappa/\kappa_{iso}$ with respect to the parameter $\mu$. Relations (\ref{Noconicald}) and (\ref{Noconicalq}) yield a prescription for  the multiple moments in terms of $\mu$, fine-tuned to ensure conical regularity, where  $\mu$ is a parameter of the background undistorted black ring.  As noted above, we have restricted our analysis to small multiple-moment values, so that the resulting fine-tuned multiple moments are likewise small, i.e., less than $-1$ [large values of $\mu$ imply correspondingly large  multiple moment values using Eqs. (\ref{Noconicald}) and (\ref{Noconicalq})].  For the dipole multiple moment, we have $\mu \in (0,0.656)$, and for the quadrupole multiple moment, we have $\mu \in (0,0.771)$. In the left diagram of Fig. \ref{kappa}, we consider the case of the dipole multiple moment by setting $a_2=0$ and using Eq. (\ref{Noconicald}) to relate $a_1$ with the parameter $\mu$:
\be
\tilde{\kappa}=e^{\frac{-\ln(1-\mu)+\ln(1+\mu)}{4\mu-4}} \, .
\ee where increasing $\mu$ implies increasingly negative $a_1$. 

\begin{center}
\begin{figure}[t]
\centering
  \includegraphics[width=7cm]{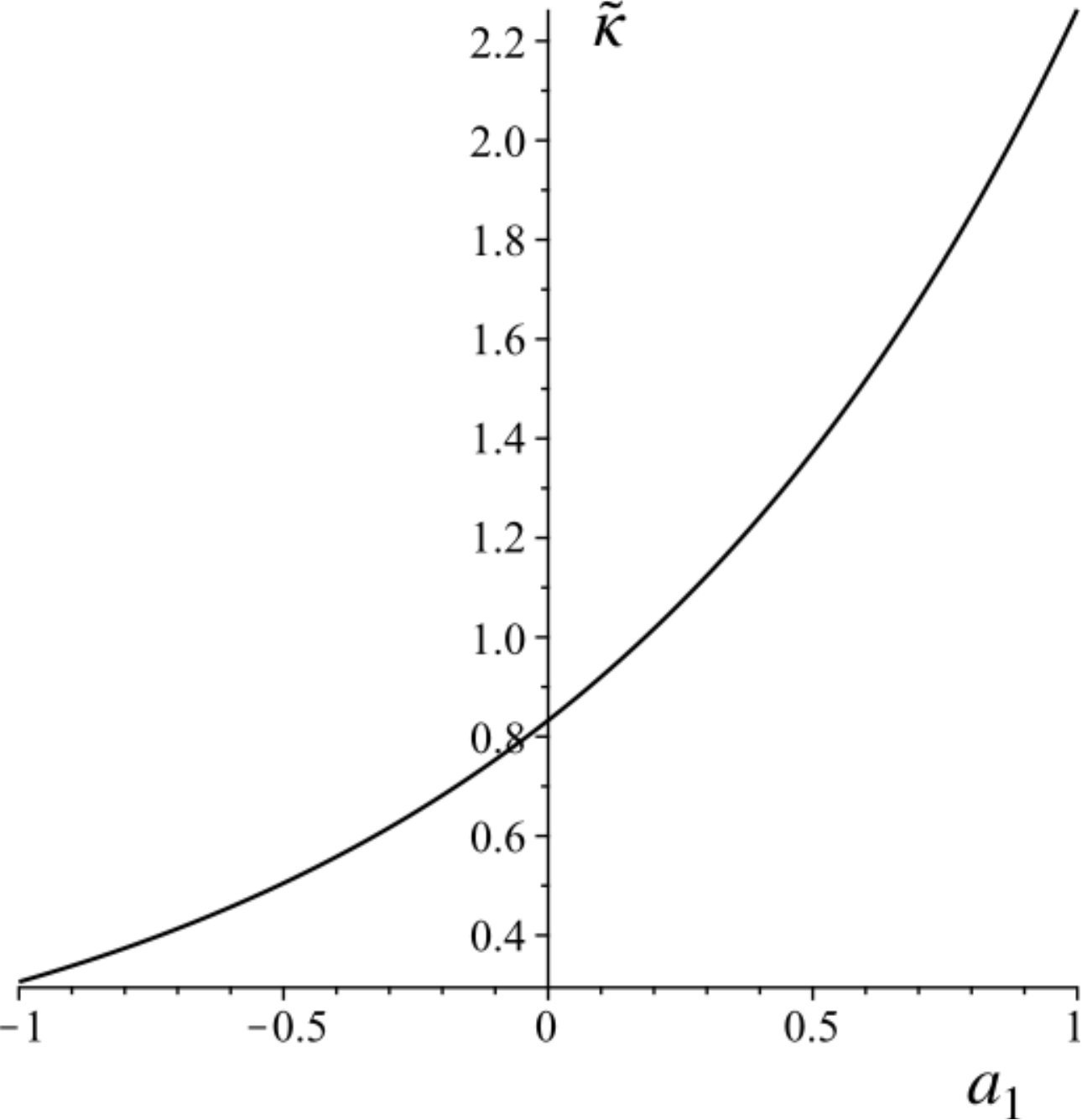}~~~~~~~
   \includegraphics[width=7cm]{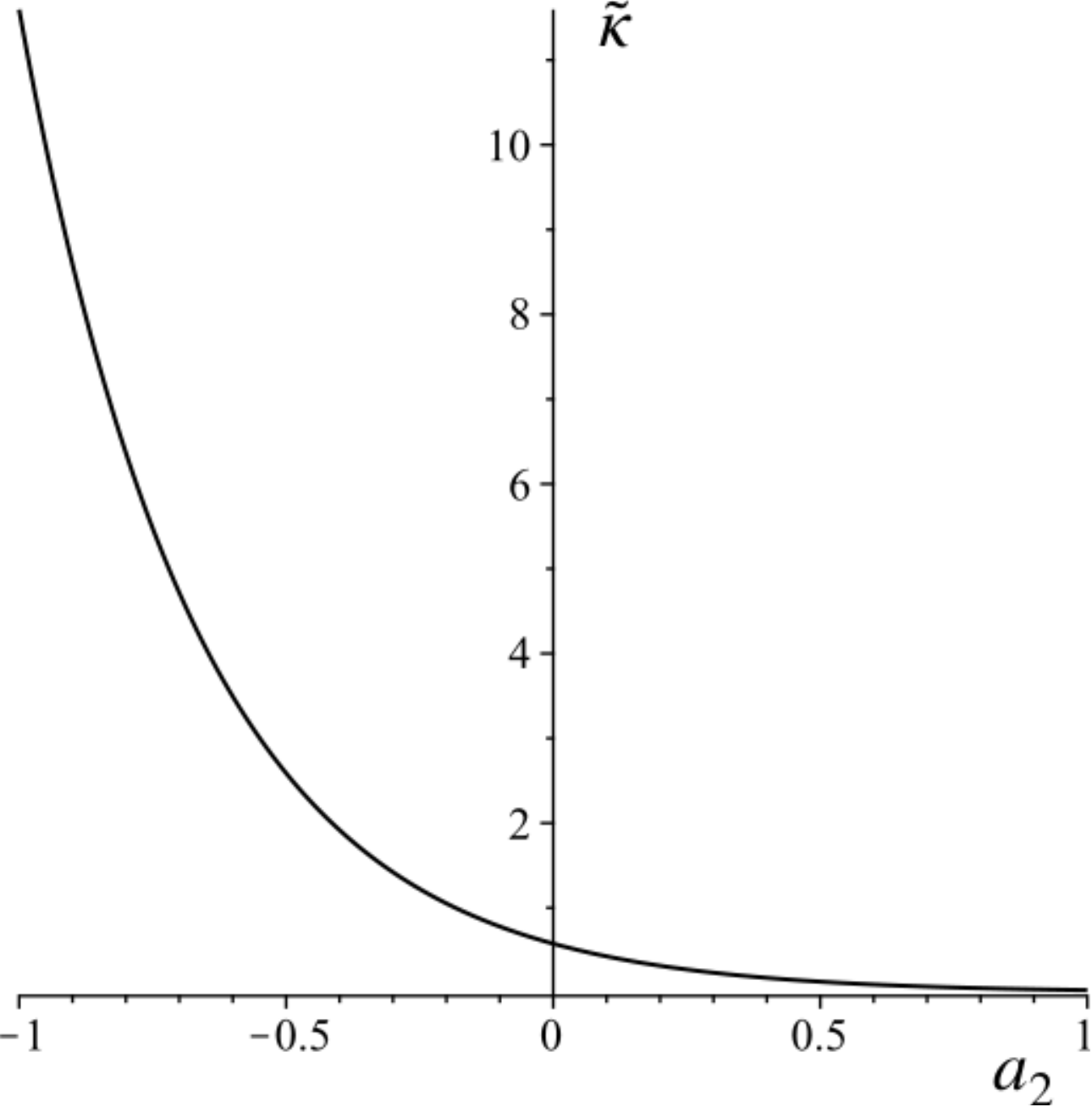}
   \caption{On the left, we have plotted the behavior of $\tilde{\kappa}$ with respect to the dipole multiple moment $a_1$ for $\mu=0.5$. On the right, we have plotted the behavior of $\tilde{\kappa}$ with respect to the quadrupole multiple moment $a_2$ for $\mu=0.5$.\label{kappa2}}
\end{figure}
\end{center}

In the right diagram of Fig. \ref{kappa}, we consider a quadrupole multiple moment by setting $a_1=0$ and using Eq. (\ref{Noconicalq}) to relate $a_2$ with the parameter $\mu$:
\be
\tilde{\kappa}=e^{\frac{[\ln(1-\mu)-\ln(1+\mu)](2\mu^2-1)}{4(\mu^2-1)}} \, ,
\ee  where again increasing $\mu$ implies increasingly negative $a_2$. We see that an increasingly large dipole moment weakens the surface gravity of the distorted black ring, whereas for an increasingly large quadrupole moment this happens only up to a certain minimal value. Beyond this value, the surface gravity of the black ring increases, ultimately becoming larger than its undistorted counterpart as the quadrupole parameter attains its maximal negative value.
 We now consider a distorted black ring having both dipole and quadrupole distortions. In  Fig. \ref{kappa2}, we have plotted the behavior of $\tilde{\kappa}=\kappa/\kappa_{iso}$ with respect to the multiple moments for a fixed parameter $\mu=0.5$ in the case of a dipole-quadrupole.  For joint dipole-quadrupole distortion, we consider two cases. On the left, we fine-tune  $a_2$, leaving $a_1$ as a free parameter. We have plotted the behavior of $\tilde{\kappa}$ with respect to the dipole multiple moment $a_1$ by replacing
\be\n{a2}
a_2 =-\frac{ \ln\left(\frac{1+\mu}{1-\mu}\right)\mu^4-3\mu^4a_1+3\mu^3a_1}{3\mu^2(1-\mu^2)}\, ,
\ee which is derived from the ``no conical singularity'' condition [Eq. (\ref{Noconical0})] for the dipole-quadrupole case. On the right,  we have fine-tuned $a_2$ and plotted the behavior of $\tilde{\kappa}$ with respect to the quadrupole multiple moment $a_2$ by inverting Eq. (\ref{a2}):
\be
a_1 =-\frac{ \ln\left(\frac{1+\mu}{1-\mu}\right)\mu^2-3\mu^2a_2+3a_2}{3\mu(1-\mu^2)}\, ,
\ee or alternatively using the conical regularity condition [Eq. (\ref{Noconical0})]. 

For any fixed $\mu$, it is straightforward to see that $a_2$ is a monotonically decreasing function of $a_1$. We find, regardless of the value of $\mu$, that the behavior of $\tilde{\kappa}$ remains the same: it monotonically increases with the dipole moment $a_1$ and thus monotonically decreases with the quadrupole moment $a_2$, as illustrated in the left and right diagrams of Fig. \ref{kappa2}, respectively.
We now calculate the Kretchmann scalar (invariant), $\mathcal{K}=R_{\alpha\beta\gamma\delta}R^{\alpha\beta\gamma\delta}$, of the spacetime on the horizon, where $R_{\alpha\beta\gamma\delta}$ is the Riemann curvature tensor. There exists is a simple
relation between the Kretschmann scalar calculated on
the horizon of a five-dimensional, static, distorted black
hole/object and the trace of the square of the Ricci tensor of its
horizon surface \cite{13s}
\be
\mathcal {K}_{\text H} = 6(\mathcal{R}_{AB}\mathcal{R}^{AB})_{\text{HS}} \, . 
\ee
where the index HS stands for the horizon surface and the index H corresponds to the horizon. It follows that the Kretschmann scalar is regular on the black hole horizon, if the distortion fields $\hu$, $\hw$, and $\hv$ are smooth on a regular horizon, which is the case according to the Eqs. (\ref{huHorizon}) and (\ref{hvHorizon}). 

\begin{center}
\begin{figure}[t]
\centering
  \includegraphics[width=7cm]{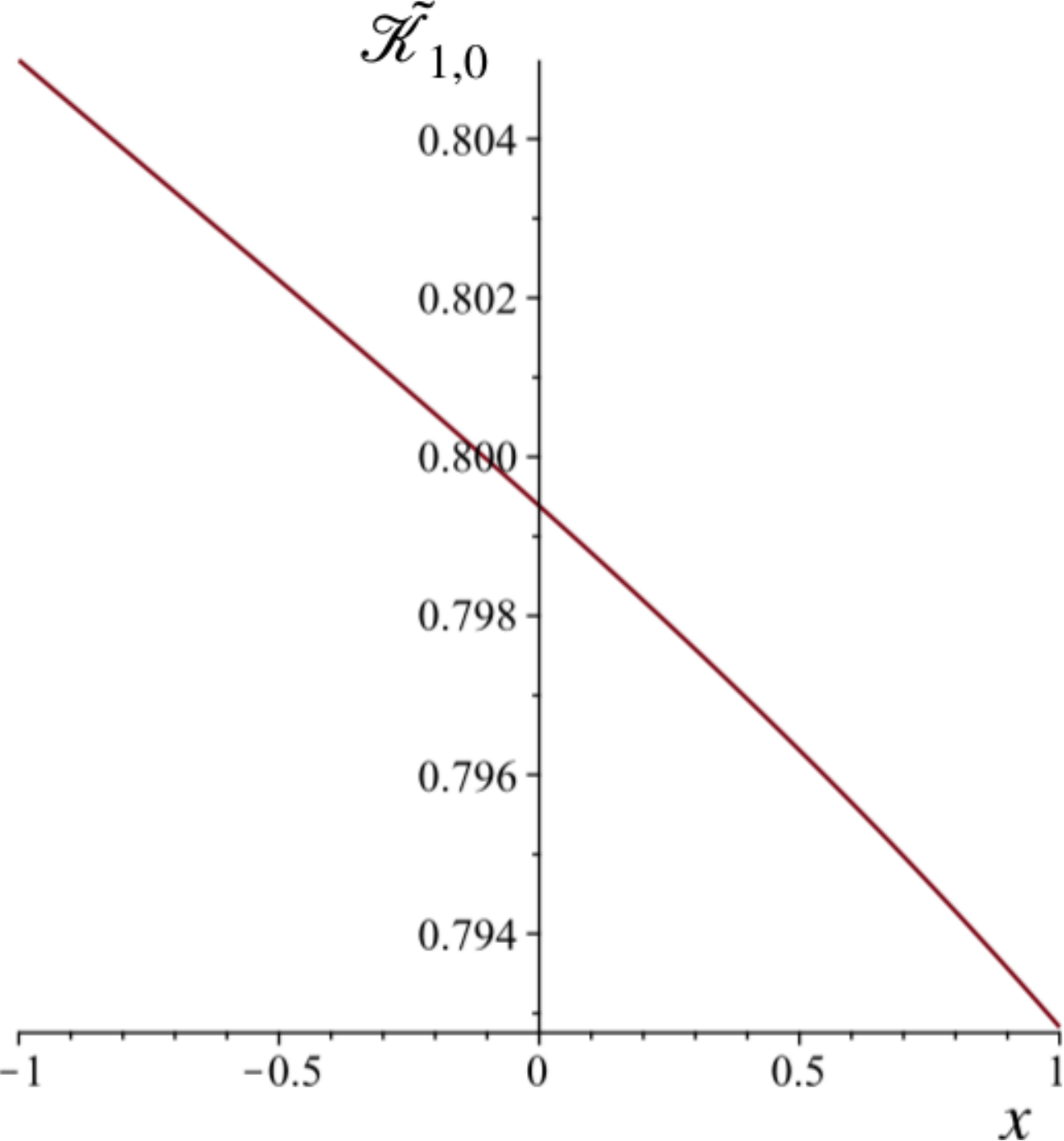}~~~~~~~
   \includegraphics[width=7cm]{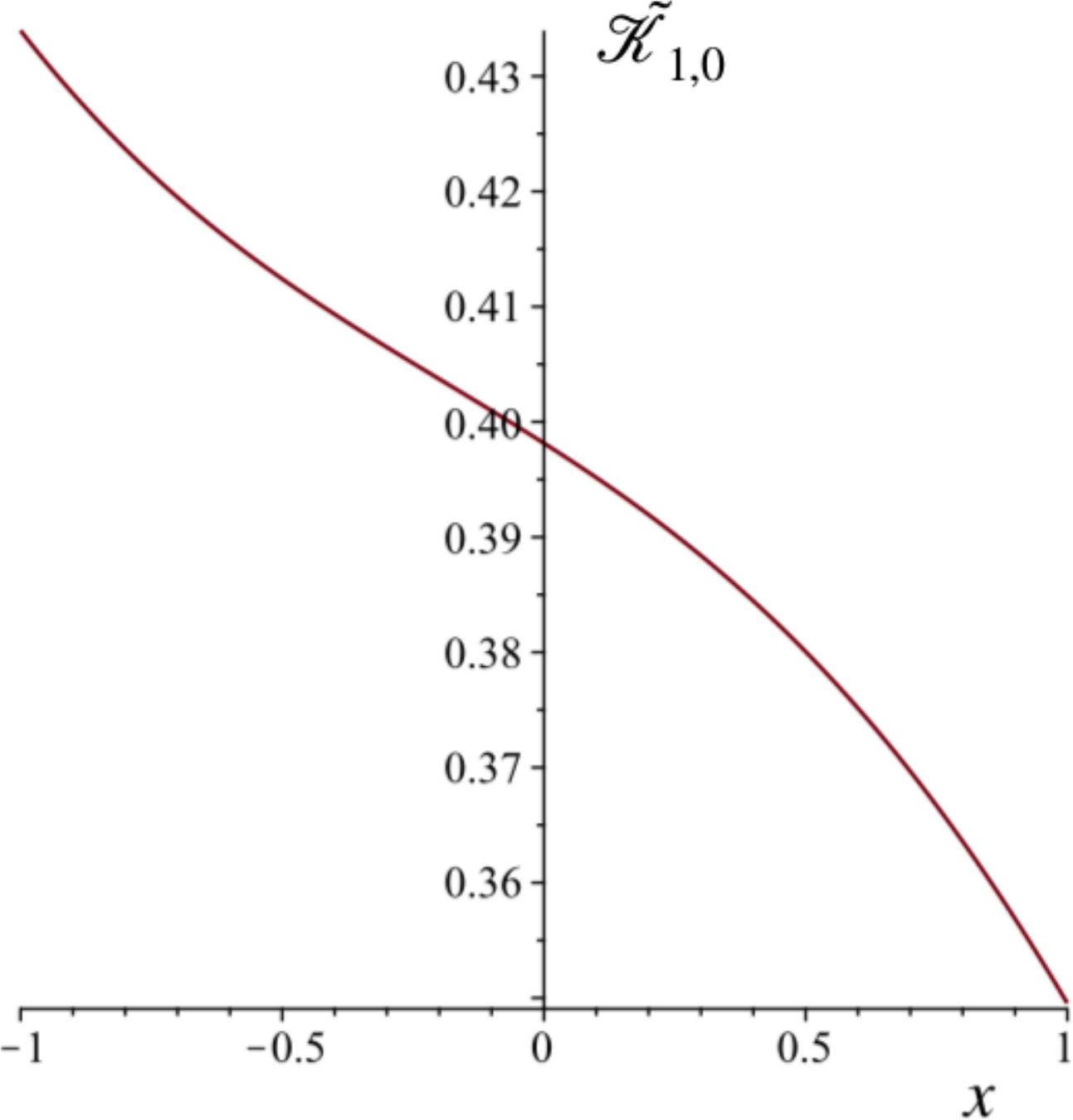}
   \caption{On the left, behavior of $\tilde{\mathcal{K}}_{1,0}$ with respect to the coordinate $x$ for a dipole multiple moment and for $\mu=0.1$. On the right, behavior of $\tilde{\mathcal{K}}_{1,0}$ with respect to the coordinate $x$ for a dipole multiple moment and for $\mu=0.3$.\label{D1}}
\end{figure}
\end{center}

\begin{center}
\begin{figure}[t]
\centering
  \includegraphics[width=7cm]{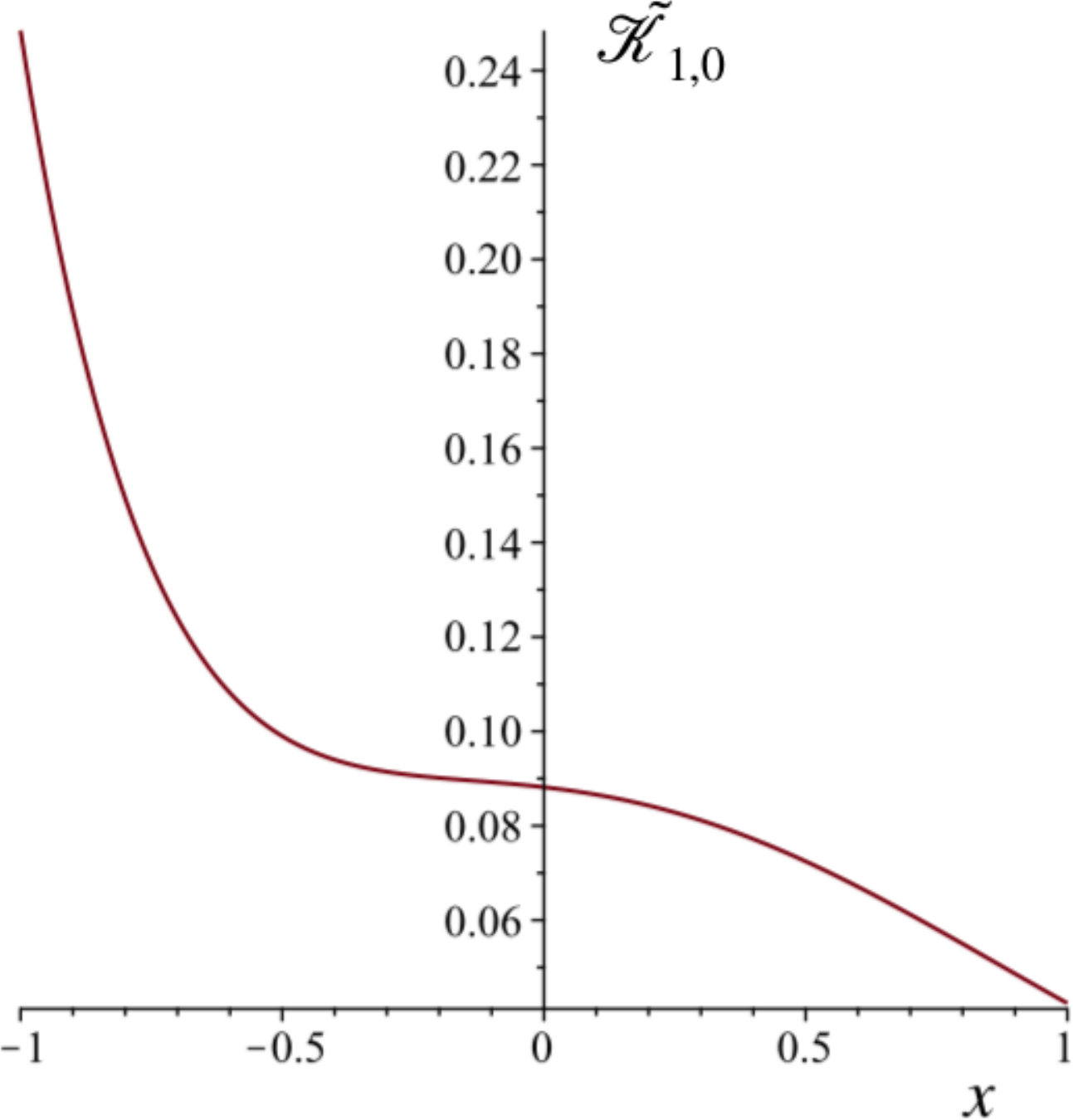}~~~~~~~
   \includegraphics[width=7cm]{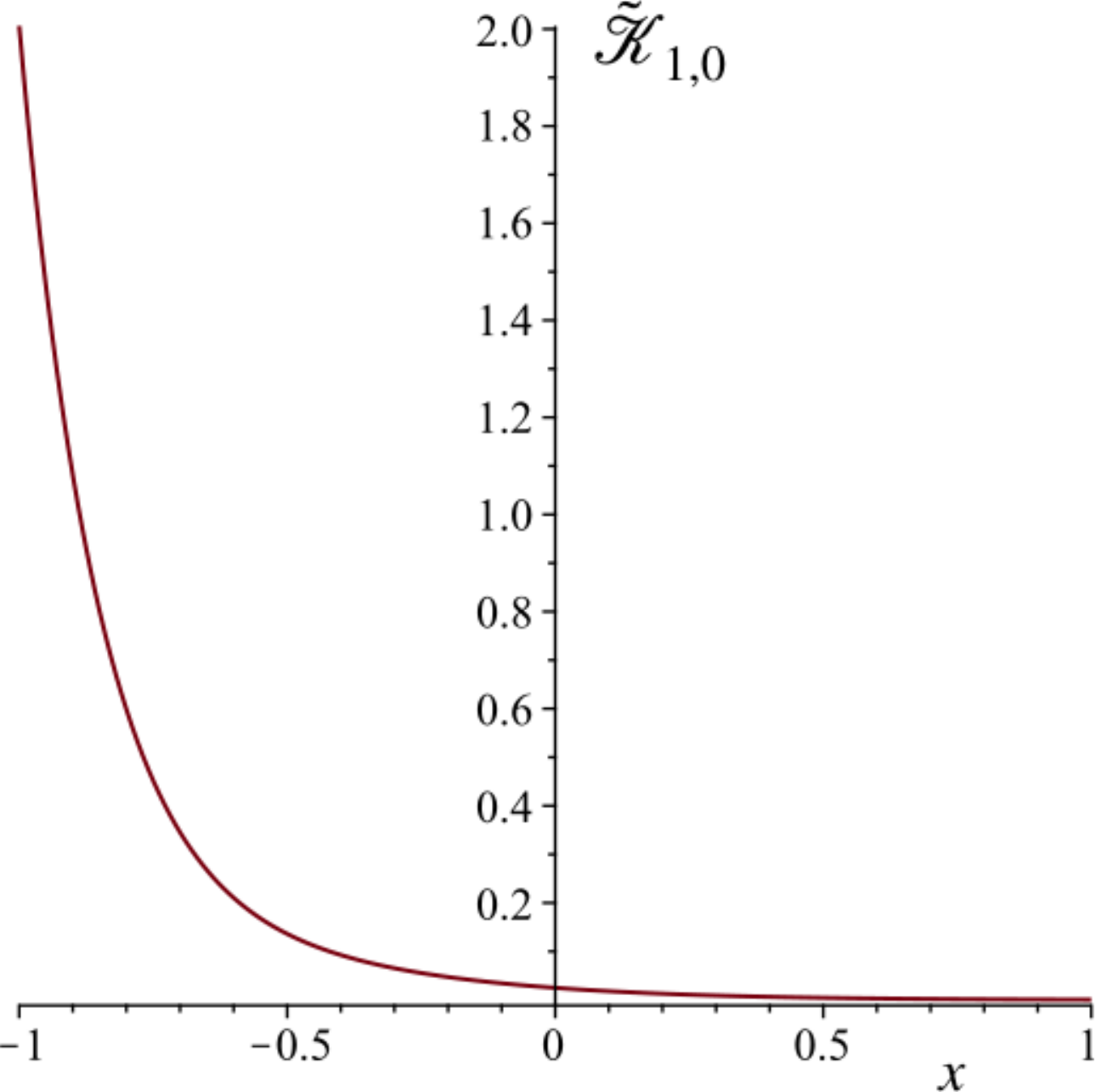}
   \caption{On the left, behavior of $\tilde{\mathcal{K}}_{1,0}$ with respect to the coordinate $x$ for a dipole multiple moment and for $\mu=0.5$. On the right, behavior of $\tilde{\mathcal{K}}_{1,0}$ with respect to the coordinate $x$ for a dipole multiple moment and for $\mu=0.65$.\label{D2}}
\end{figure}
\end{center}

\begin{center}
\begin{figure}[t]
\centering
  \includegraphics[width=7cm]{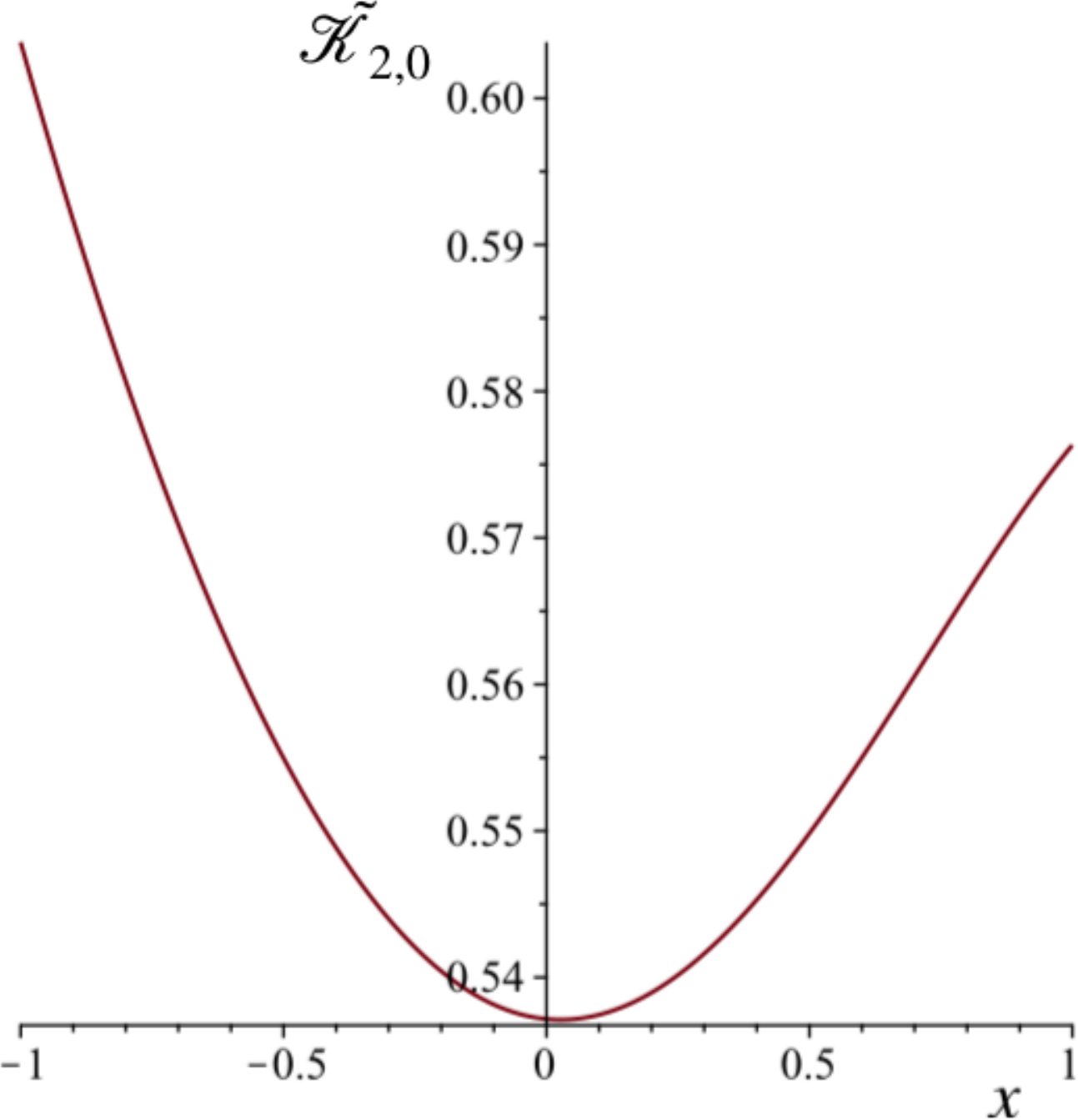}~~~~~~~
   \includegraphics[width=7cm]{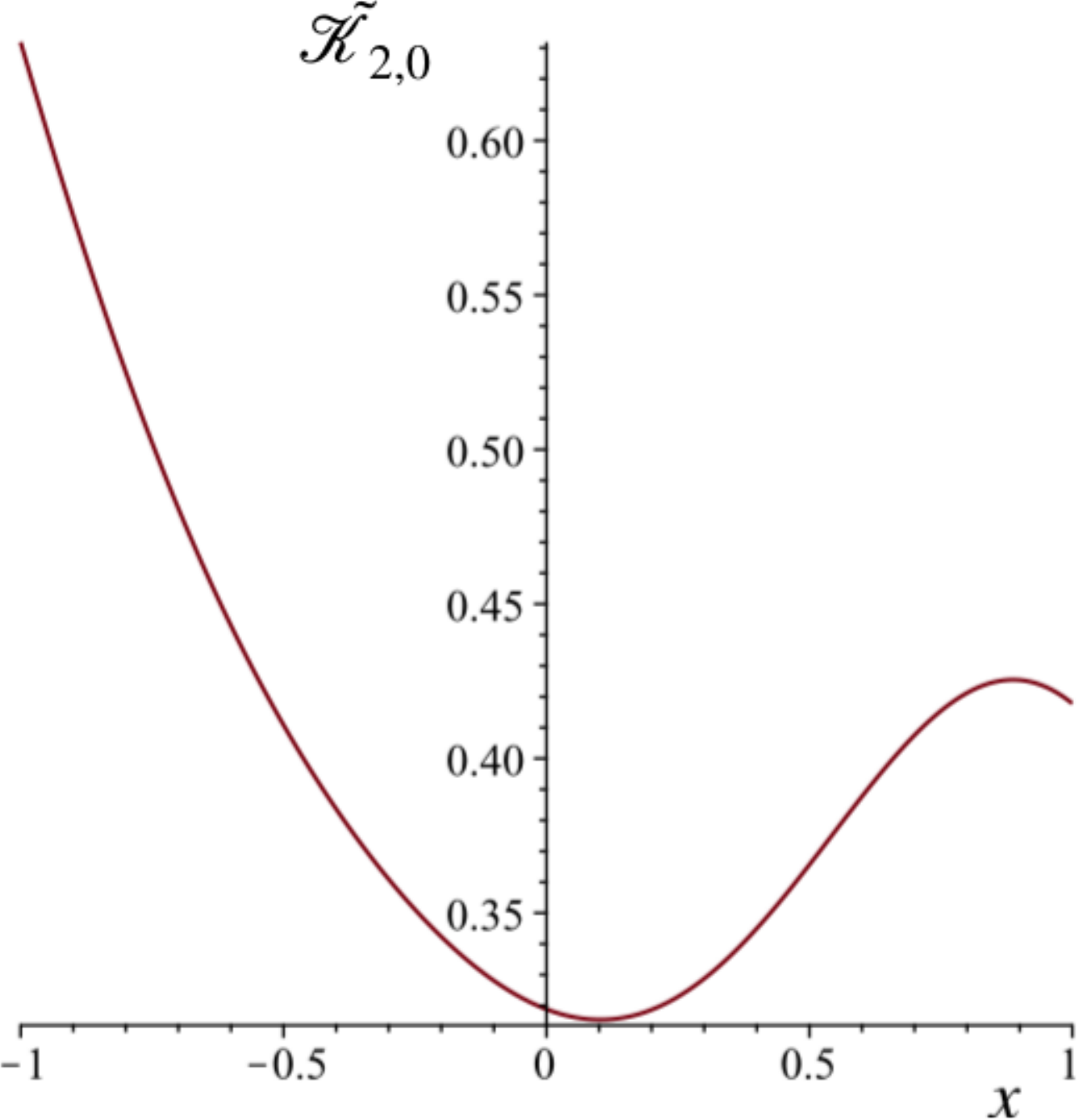}
   \caption{On the left, behavior of $\tilde{\mathcal{K}}_{2,0}$ with respect to the coordinate $x$ for a quadrupole multiple moment and for $\mu=0.3$. On the right, behavior of $\tilde{\mathcal{K}}_{2,0}$ with respect to the coordinate $x$ for a quadrupole multiple moment and for $\mu=0.5$.\label{Q1}}
\end{figure}
\end{center}

\begin{center}
\begin{figure}[t]
\centering
  \includegraphics[width=7cm]{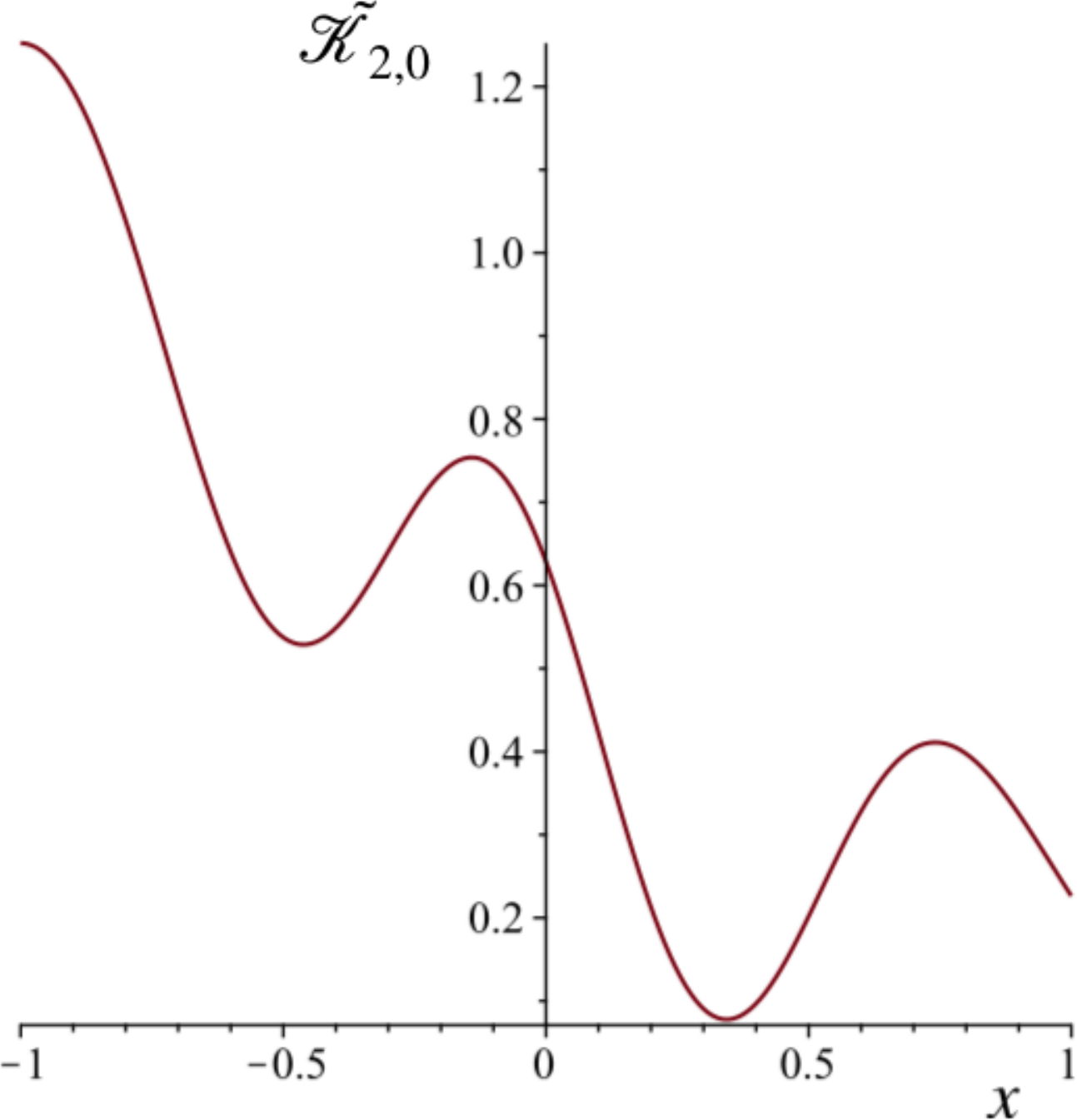}~~~~~~~
   \includegraphics[width=7cm]{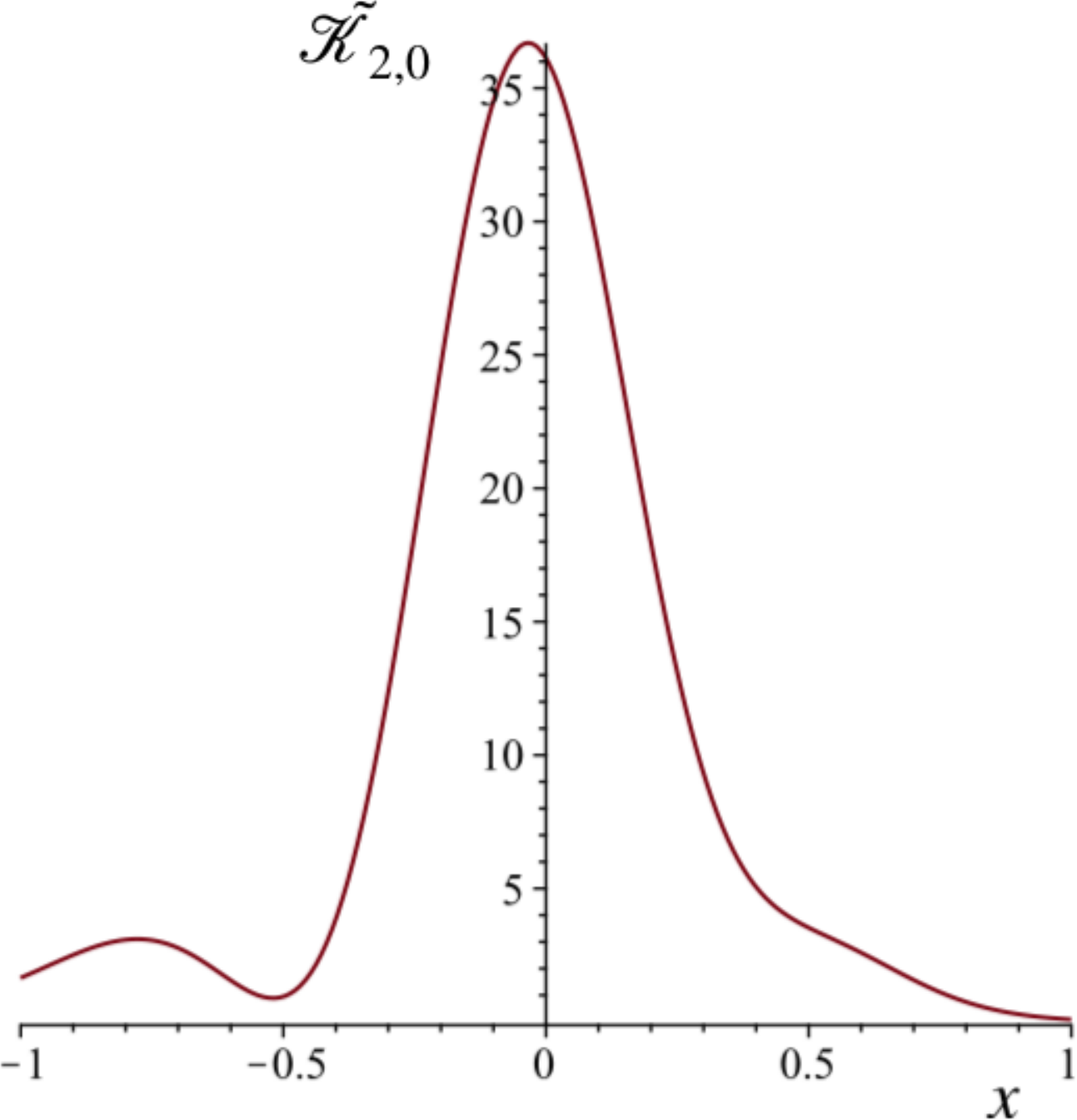}
   \caption{On the left, behavior of $\tilde{\mathcal{K}}_{2,0}$ with respect to the coordinate $x$ for a quadrupole multiple moment and for $\mu=0.7$. On the right, behavior of $\tilde{\mathcal{K}}_{2,0}$ with respect to the coordinate $x$ for a quadrupole multiple moment and for $\mu=0.8$.\label{Q2}}
\end{figure}
\end{center}

The Kretschmann scalar calculated on the horizon of the undistorted black ring is given by 
\be
\mathcal{K}|_H=\frac{A^4(3\mu^4x^4-2\mu^4x^2-12\mu^3x^3+3\mu^4+4\mu^3x+20\mu^2x^2-8\mu^2-16\mu x+8)}{4\mu^4(1-\mu x)^4}\, ,
\ee
and the ratio of this quantity to that of a black ring with a monopole distortion is $\exp(4b_0+4a_0)$. The corresponding ratio for a dipole distortion to a monopole distortion is 
\ba
\tilde{\mathcal{K}}_{1,0}=&&e^{\frac{(4\mu x+3)a_1}{\mu}}\biggl[1+\frac{(\mu x-1)}{(6x^4-4x^2+6)\mu^4+(-24x^3+8x)\mu^3+(40x^2-16)\mu^2-32\mu x+16}\biggl(7(\mu x-1)^3(x-1)^2(x+1)^2a_1^4\nonumber\\
&&+(\mu x-1)^2 (19\mu x^2-21 \mu+2x)(x^2-1)a_1^3\nonumber\\
&&+(\mu x-1)(41 \mu^2 x^4-16 \mu^2 x^2-70\mu x^3+19 \mu^2-18 \mu x+48 x^2-4) a_1^2\nonumber\\
&&-4\left((x^4-3x^2-2)\mu^3-2(3x^3-x)\mu^2+5(3x^2+1)\mu-12x\right)a_1\biggl)\biggl] \, . 
\ea  
whereas 
\ba
\tilde{\mathcal{K}}_{2,0}=&&e^{\frac{(4\mu^2x^2-6\mu^2+3)a_2}{\mu^2}}\biggl[1+\frac{(\mu x-1)}{(3x^4-2x^2+3)\mu^4+(-12x^3+4x)\mu^3+(20x^2-8)\mu^2-16\mu x+8}\biggl(\nonumber\\
&&56x^4(\mu x-1)^3(x-1)^2(x+1)^2a_2^4\nonumber\\
&&+4(\mu x-1)^2x^2(23\mu x^3-25\mu x-2x^2+4)(x^2-1)a_2^3\nonumber\\
&&+2(\mu x-1)(91\mu^2x^6-72\mu^2x^4-176\mu x^5+25\mu^2x^2+106\mu x^3+104x^4-18\mu x-72x^2+12) a_2^2\nonumber\\
&&-12\left(\mu^3x^5-3x^3\mu^3+\frac{2}{3}\mu^3x-4\mu^2x^4+4x^2\mu^2-\frac{4}{3}\mu^2+\frac{25}{3}\mu x^3-\frac{5}{3}\mu x-6 x^2+2\right)a_2\biggl)\biggl]
\ea
is the corresponding ratio for a quadrupole distortion to a monopole one. 
In Figs. \ref{D1} and \ref{D2}, we have plotted the behavior of $\tilde{\mathcal{K}_{1,0}}$ with respect to coordinate $x$ for a dipole multiple moment and for four different values of the parameter $\mu$. In each case, the value of the dipole multiple moment is chosen such that there is no conical singularity in the black ring. 
In Figs. \ref{Q1} and \ref{Q2}, we have plotted the behavior of $\tilde{\mathcal{K}_{2,0}}$ with respect to the coordinate $x$ for a quadrupole multiple moment and for four different values of the parameter $\mu$. In each case, the value of the quadrupole multiple moment is chosen such that there is no conical singularity in the black ring.

\section{Summary}

 We have constructed the metric of a distorted five-dimensional static black ring. The solution represents a local black ring distorted by external static and neutral distributions of matter sources.   One of the interesting features of this local solution is that by careful fine-tuning of the external distorting sources we can have a solution that is free of the conical singularity that is present in the isolated black ring. 
 
Our analysis of the Kretschmann scalar on the horizon of the distorted black ring indicates that these distortions do not produce any new singularities, since the distortion fields $\hu$, $\hw$ and $\hv$ are regular on the horizon. We have also analyzed the behavior of the distorted black ring's surface gravity. For joint dipole-quadrupole distortion we  found that increasingly large (negative) quadrupole distortions weaken the surface gravity of the black ring, and increasingly large (positive) dipole distortions strengthen the surface gravity of the black ring. However, for quadrupole distortions increasingly large (negative) dipole distortions strengthen the surface gravity of the black ring, and increasingly large (positive) quadrupole distortions weaken the surface gravity of the black ring. The surface gravity of the distorted black hole can become as much as 10 times larger than the undistorted black ring.

We have also analyzed the behavior of the Kretschmann scalar of the horizon for the dipole and quadrupole distortions. For the dipole distortion and $a_1$ greater than $-0.751$ (which is an appropriate value of $a_1$ for a black ring with $\mu=0.612$), the effect of the dipole distortion is to decrease the curvature of the horizon. For $a_1$ less than this value, the Kretschmann scalar of the horizon can increase under the effect of distortion, with its maximum value at $x=-1$. For the quadrupole distortion, and $a_2$ greater than $-0.37$ (which is an appropriate value of $a_2$ for a black ring with $\mu=0.612$), the effect of the quadrupole distortion is to decrease the curvature of the horizon. However, for $\mu>0.612$, the external sources can increase the curvature of the horizon. 

For future work, it would be interesting to study the effects of distortions on the shape of the horizon and the stretched singularity of the black ring, and investigate whether certain duality transformations exist  between the horizon and stretched singularity, similar to that   observed  for  distorted four- and five-dimensional Schwarzschild black holes \cite{Frolov:2007xi,13s}. 

\section*{Acknowledgements} S.A. appreciated the hospitality of the KITP, as a KITP scholar. This research was supported in part by the National Science Foundation under Grant No. NSF PHY-1748958  and by the Natural Sciences and Engineering Research Council of Canada. 



\end{document}